\documentclass[useAMS,usenatbib,usegraphicx]{mn2e}

\usepackage{subfigure}
\usepackage{url}
\usepackage{comment}
\usepackage{color}
\usepackage{dcolumn}
\usepackage{enumitem}
\usepackage{ulem}

\newcommand*{\rom}[1]{\uppercase\expandafter{\romannumeral #1}}
\newcommand\ion[2]{#1$\;${\small\rmfamily\rom{#2}}\relax}%

\newcommand\aap{{A\&A}}%
\newcommand\araa{{ARA\&A}}%
\newcommand\aaps{{A\&AS}}%
\newcommand\mnras{{MNRAS}}%
\newcommand\apj{{ApJ}}%
\newcommand\apjs{{ApJS}}%
\newcommand\apjl{{ApJ}}%
\newcommand\aj{{AJ}}%
\newcommand\pasp{{PASP}}%
\setlist[enumerate]{noitemsep}
\setlist[enumerate,1]{leftmargin=*}
\setlist[itemize]{noitemsep}
\setlist[itemize,1]{leftmargin=*}
\setlist[description]{noitemsep}
\setlist[description,1]{leftmargin=*}
\setenumerate[0]{label=(\arabic*)}

\makeatother

\title[Abundances of NGC~1718 stars]{Chemical Abundances of Two Stars in the Large Magellanic Cloud Globular Cluster NGC~1718}

\author[Sakari et al.]{Charli
M. Sakari$^{1}$\thanks{E-mail: sakaricm@u.washington.edu}, Andrew
McWilliam$^{2}$, and George Wallerstein$^{1}$\\
$^{1}$ Department of Astronomy, University of Washington, Seattle, WA
98195-1580, USA\\
$^{2}$ Observatories of the Carnegie Institute of Washington,
Pasadena, CA, USA\\
}

\begin{document}

\maketitle

\label{firstpage}

\begin{abstract}
Detailed chemical abundances of two stars in the intermediate-age
Large Magellanic Cloud (LMC) globular cluster NGC~1718 are presented,
based on high resolution spectroscopic observations with the MIKE
spectrograph.  The detailed abundances confirm NGC~1718 to be a fairly
metal-rich cluster, with an average $[\rm{Fe/H}] \sim -0.55\pm0.01$.
The two red giants appear to have primordial O, Na, Mg, and Al
abundances, with no convincing signs of a composition difference
between the two stars---hence, based on these two stars, NGC~1718
shows no evidence for hosting multiple populations.  The Mg abundance
is lower than Milky Way field stars, but is similar to LMC field stars
at the same metallicity. The previous claims of very low [Mg/Fe] in
NGC~1718 are therefore not supported in this study.  Other abundances
(Si, Ca, Ti, V, Mn, Ni, Cu, Rb, Y, Zr, La, and Eu) all follow the LMC
field star trend, demonstrating yet again that (for most elements)
globular clusters trace the abundances of their host galaxy's field
stars.  Similar to the field stars, NGC~1718 is found to be mildly
deficient in explosive $\alpha$-elements, but moderately to strongly
deficient in O, Na, Mg, Al, and Cu, elements which form during
hydrostatic burning in massive stars.  NGC~1718 is also enhanced in
La, suggesting that it was enriched in ejecta from metal-poor AGB
stars.
\end{abstract}


\section{Introduction}\label{sec:Intro}
Detailed abundances of stars in globular clusters (GCs) are essential
for two primary goals: 1) understanding the nature of GC formation
(e.g., \citealt{Gratton2012}) and 2) tracing the properties of field
star populations in distant, unresolved galaxies (e.g.,
\citealt{Colucci2013}, \citealt{Sakari2015}).  For most elements, the
abundances of GC stars trace those of the field stars in their birth
environment \citep{Pritzl2005,Hendricks2016}, providing probes of a
galaxy's star formation history, abundance gradients, chemical
evolution, and assembly history.  However, for a handful of other
elements, from light elements like C, N, O, and Na to heavy neutron
capture elements like Ba and Eu, Milky Way (MW) GCs host star-to-star
variations that are unique to GCs and are not seen in most field stars
(detections of field stars with these abundance variations are thought
to be accreted from dissolved GCs; e.g., \citealt{Martell2016}).
Similar variations have also been observed in classical, old GCs in
dwarf galaxies, including the Large Magellanic Cloud (LMC;
\citealt{Johnson2006,Mucciarelli2009,Mucciarelli2010}).  These
variations seem to be present in M31 GCs as well, and do affect the
integrated abundances from distant, unresolved GCs
(\citealt{Colucci2014}, \citealt{Sakari2013,Sakari2016}).  Despite the
prevalence of GC multiple populations, the cause of these abundance
variations is not yet well-understood.  Observations of GCs outside of
the MW, particularly ones that are unlike standard MW GCs, are
necessary to understand GC formation.  Without a more complete
understanding of the multiple populations in GCs, interpreting
integrated abundances of unresolved clusters remains difficult.

One cluster that is particularly intriguing for detailed abundance
studies is the intermediate-age LMC cluster
NGC~1718. {\it Hubble Space Telescope} photometry has revealed that
the cluster is of intermediate age and moderately high metallicity,
with an age $\sim2$ Gyr and $[\rm{Fe/H}]\sim-0.4$
\citep{Brocato2001,Kerber2007}.  Calcium triplet spectroscopy of three
cluster stars suggests a slightly lower value of $[\rm{Fe/H}]\sim-0.8$
\citep{Grocholski2006}, while comparisons with other LMC clusters of a
similar age suggest that NGC~1718 should have a metallicity of
$[\rm{Fe/H}]\sim-0.42$ \citep{MackeyGilmore2003}.   NGC~1718 is far
too massive \citep{Baumgardt2013} to be an open cluster---it is
therefore distinctly different from the classical, metal-rich MW GCs,
which are all older than 10 Gyr.\footnote{Though there are metal-rich,
  intermediate-age MW clusters (e.g. Palomar~1; \citealt{Sakari2011}),
  these clusters are much less massive than NGC~1718, and are not
  obviously classical GCs.}  Intermediate-age GCs like NGC~1718 are
therefore excellent targets for studying the nature of GC multiple
populations and for examining relatively recent star formation in the
LMC.

The first detailed abundances for NGC~1718 were derived by
\citet{Colucci2011,Colucci2012} from integrated light (IL)
spectroscopy, where a single spectrum is obtained from the entire
stellar population.  With a spectrum that only covered $\sim23$\% of
the cluster, they found NGC~1718 to be a moderate-metallicity
($[\rm{Fe/H}] = -0.7$), intermediate-age (1.25-2 Gyr), solar [Ca/Fe]
cluster with a very low Mg abundance ($[\rm{Mg/Fe}] = -0.9$).  While
Mg is not always expected to trace the heavier $\alpha$-elements like
Ca and Ti in a low mass galaxy, it is difficult to explain such a low
Mg abundance through canonical chemical evolution scenarios.  Colucci
et al. suggested that NGC~1718's low integrated Mg abundance indicated
that the cluster was enriched solely through ejecta from a Type Ia
supernova, a scenario which would increase Fe significantly without
appreciably changing Mg \citep{TsujimotoBekki2012}. However, similarly
low-[Mg/Fe] field stars have not been found in the LMC
\citep{Lapenna2012}, as would be expected if NGC~1718 formed in a
environment enriched only by a Type Ia supernova.

Because of its age and mass, NGC~1718 is also interesting as a GC.
The source of multiple populations within GCs continues to be debated;
a particularly contentious point is whether the multiple populations
are actually multiple {\it generations} with a very small ($\sim 100$
Myr) age spread.  Under many multiple population formation scenarios,
the ratios of ``primordial,'' first generation stars (with normal
abundance ratios) to ``extreme'' second generation stars (with
enhanced [Na/Fe] and deficient [O/Fe]) requires a significant amount
of mass loss from the first generation prior to the formation of the
second generation.  Convincing age spreads have not yet been detected
in any of the younger LMC GCs, including NGC~1718. (Despite its broad
main sequence, which suggests that the cluster hosts an age spread,
its red clump implies that NGC 1718 does host a population with a
single age; \citealt{Niederhofer2016}.)  However, although multiple
populations have been spectroscopically confirmed in LMC and Small
Magellanic Cloud GCs older than $\sim 8$ Gyr
\citep{Johnson2006,Mucciarelli2009,Mucciarelli2010,Hollyhead2016} 
{\it there is not yet any convincing evidence for the existence of
  multiple populations in GCs younger than $\sim 8$ Gyr}
\citep{Mucciarelli2008,Mucciarelli2011,Mucciarelli2014}.  This
difference implies that the LMC GCs that formed more recently may be
fundamentally different from the old GCs that formed early in
the universe.  The intermediate-age and young LMC GCs are therefore
important targets for understanding the nature of multiple populations
within GCs.

This paper presents the first abundance analyses of individual stars
in NGC~1718, from high resolution spectroscopy of two cluster
members.  These abundances are calculated differentially with respect
to the well studied giant Arcturus, and therefore have relatively low
systematic errors.  The detailed abundances of the two stars are then
examined in light of NGC~1718's context as a young GC and as a member
of the LMC.

\section{Observations and Data Reduction}\label{sec:Observations}

\begin{table*}
\centering
\hspace*{-0.5in}
\begin{minipage}{165mm}
\begin{center}
\caption{Target information.\label{table:Targets}}
  \begin{tabular}{@{}lcccccccccccc@{}}
  \hline
 & 2MASS IDs & RA (hms) & Dec (dms) & $J$ & $K$ & Observation & $t_{\rm{exp}}$ & Seeing & S/N & $v_{\rm{helio}}$ \\
 &  & J2000 & J2000 &  &  & Dates & (sec) & (arcsec) & $^{a}$ & (km s$^{-1}$) \\
\hline
NGC 1718-9  & 04522589-6702590 & 4:52:26.0   & $-$67:02:59.1 & 13.52 & 12.56 & 26 Feb, 2012 & 11,400 & 0.65 & 49 & $282.1\pm 1.0$\\
NGC 1718-26 & 04521682-6703242 & 4:52:16.8 & $-$67:03:24.3 & 13.95 & 12.90 & 27 Feb, 2012 & 10,800 & 0.60 & 52 & $283.4\pm 1.0$\\
\hline
\end{tabular}
\end{center}
\end{minipage}\\
\medskip
\raggedright $^{a}$ S/N is per final extracted wavelength pixel at the
peak of the H$\alpha$ order; weak lines typically have FWHM$\sim 5$
pixels.\\
\end{table*}

\begin{figure*}
\begin{center}
\centering
\subfigure[2MASS $K$-band Image]{\includegraphics[scale=0.4,trim=0.0in 0 0in 0.0in,clip]{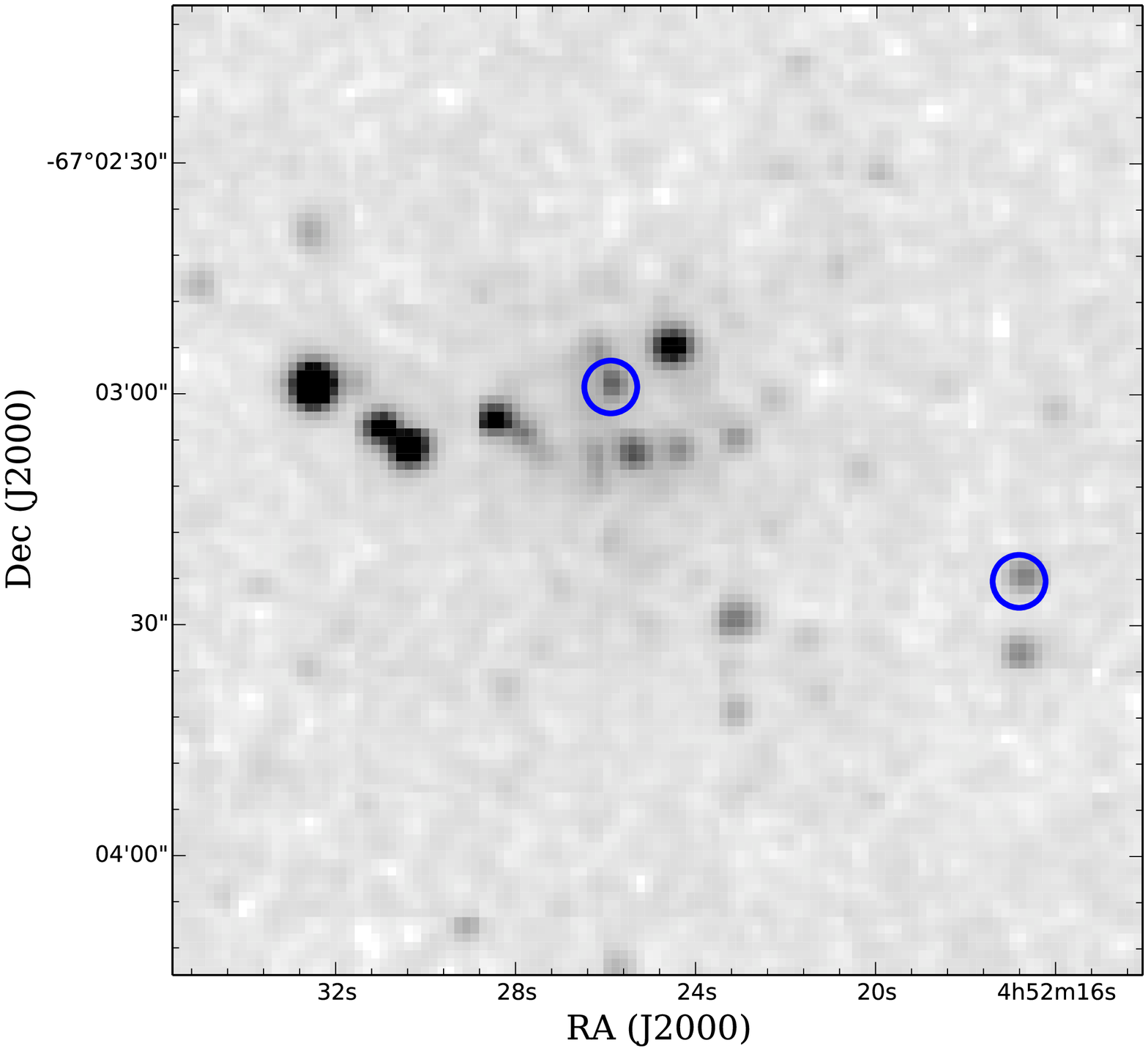}}
\subfigure[Colour-magnitude diagram]{\includegraphics[scale=0.5,trim=0.in 0 0.2in 0.5in,clip]{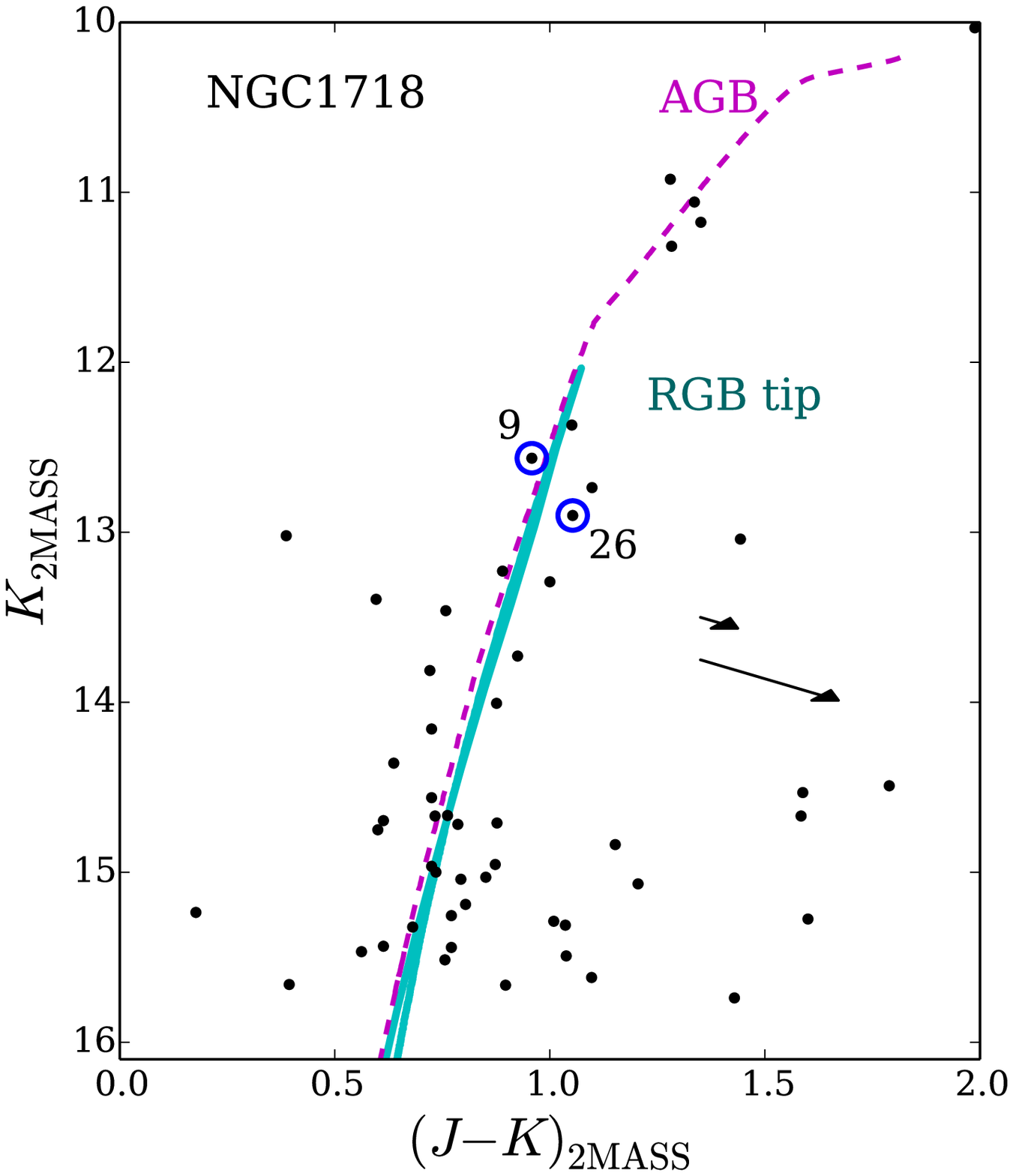}\label{fig:CMD}}
\caption{2MASS data for NGC~1718.  {\it Left: } $K$-band image with
  targets circled.  North is up and east is to the right.  {\it
    Right:} $K$ vs. $(J-K)$ colour-magnitude diagram, utilizing
  observed colours.  The two targets for the abundance analysis are 
  circled.  A 2 Gyr, z=0.004, solar-scaled
  BaSTI isochrone \citep{BaSTIref} is also shown to highlight the RGB
  (solid thick cyan line) and the AGB (magenta dashed line; the
  isochrone has an extended AGB and a mass loss parameter of $\eta =
  -0.2$ ).  Two reddening vectors are shown: one with the
  \citet{Kerber2007} value of $E(B-V)=0.1$, the other with the
  \citet{SchlaflyFinkbeiner2011} value of $E(B-V)=0.598$; the higher
  value predicts temperatures that are incompatible with the
  spectroscopic temperatures.}\label{fig-2masscmd}

\end{center}
\end{figure*}

\begin{figure*}
\begin{center}
\centering
\hspace*{-0.5in}
\includegraphics[scale=0.55,trim=0.0in 0 0in 0.0in,clip]{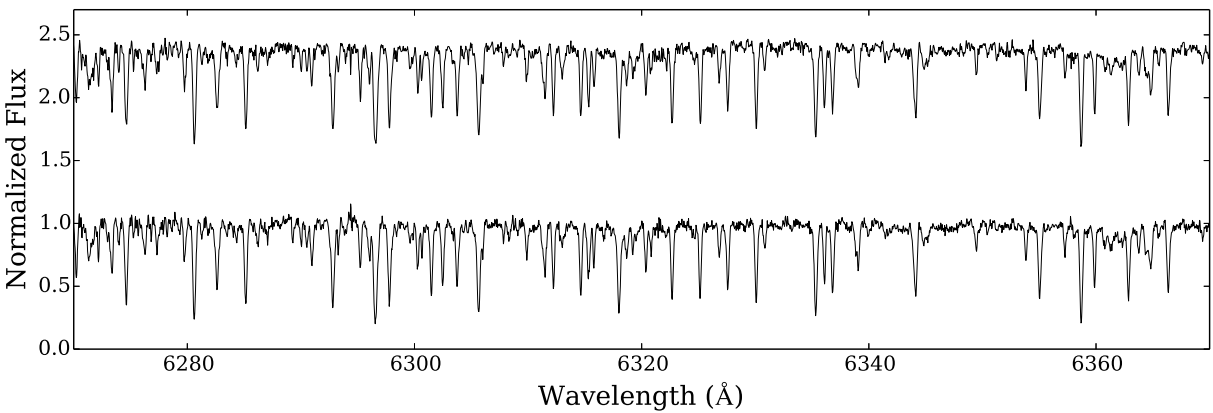}
\caption{Sample spectra in the $6270-6370$ \AA \hspace{0.025in} range.}\label{fig:Spectra}
\end{center}
\end{figure*}

Probable RGB members of NGC~1718 were selected using photometry from
the Two Micron All-Sky Survey (2MASS; \citealt{2MASSref}).
Figure~\ref{fig-2masscmd} shows a 2MASS $K$-band image of the cluster
and the $K$ versus $J-K$ colour-magnitude diagram (CMD) for stars
within 60$\arcsec$ of the cluster centre. The two targets in this
paper are identified.  Following the $\sim$2 Gyr age estimated by
\citet{ElsonFall1988} and \citet{Kerber2007}, a 2 Gyr, z=0.004 BaSTI
isochrone \citep{BaSTIref} is also displayed in
Figure~\ref{fig-2masscmd} to distinguish the RGB from the brighter AGB
stars.  The star identification numbers used here and shown in
Table \ref{table:Targets} and Figure~\ref{fig-2masscmd} were assigned
ad-hoc (by distance from input cluster centre in the 2MASS catalog).

High-resolution (R$\sim$45,000) spectra of the two target stars were
obtained on 26 and 27 February 2012, using the MIKE echelle
spectrograph, with a 0.5$\arcsec$ slit, on the Magellan/Clay
telescope.  The average atmospheric seeing was 0.6\arcsec, FWHM, on
both nights.  The usable wavelength coverage is from $\sim 5300$ to
9000 \AA.

Extraction of the spectra from the CCD data was performed using the MIKE
pipeline software from \citet{Kelson2003}.  However, subsequent
analysis employed the suite of routines from the Image Reduction and
Analysis Facility program (IRAF).\footnote{IRAF is distributed by the
  National Optical Astronomy Observatory, which is operated by the
  Association of Universities for Research in Astronomy, Inc., under
  cooperative agreement with the National Science Foundation.}  S/N
ratios at the peak of the H$_{\alpha}$ order are estimated at 49 and
52 for star \#9 and \#26, respectively, per extracted wavelength
pixel.  Typical weak stellar lines have FWHM $\sim 5$ pixels.

In order to facilitate continuum placement and EW measurement, the
spectral modulation resulting from the echelle blaze was removed by
dividing by a high S/N blaze function spectrum, which was found by
fitting the continuum  flux of the bright, extremely metal-poor, RGB
star HD~126587.   Radial velocities were determined through
cross-correlations with a high resolution, high S/N Arcturus spectrum
from
\citet{Hinkle2003}.\footnote{\url{ftp://ftp.noao.edu/catalogs/arcturusatlas/}}
The final, heliocentric radial velocities are shown in Table
\ref{table:Targets} and are in agreement with the radial velocities of
other confirmed cluster members from \citet{Grocholski2006}.
Spectra in the $6270-6370$ \AA\hspace{0.025in} range are shown in
Figure \ref{fig:Spectra}.

\section{Atmospheric Parameters}\label{sec:AtmParams}
Kurucz atmospheres\footnote{\url{http://kurucz.harvard.edu/grids.html}}
\citep{KuruczModelAtmRef} are adopted for this analysis, with an
interpolation scheme to select $T_{\rm{eff}}$ and $\log g$ values that
fall between the grid points.  Solar-scaled (ODFNEW) atmospheres are
adopted for the NGC~1718 stars, since the [$\alpha$/Fe] ratios in
these targets are low (see Section \ref{sec:Abunds}).

Photometric effective temperatures were derived utilizing the observed
2MASS $(J-K)$ colours (see Figure \ref{fig:CMD}) and the empirical
relation of \citet{GonzalezHernandezBonifacio2009}.   The reddening
and distance modulus from \citet{Kerber2007}, $E(B-V) = 0.10$ and
$(m-M)_V = 18.73$, are adopted, and the reddening relations from
\citet{McCall2004} are used to deredden the observed 2MASS
colours. These photometric temperatures are shown in Table
\ref{table:AtmParams}.  Note that the \citet{SchlaflyFinkbeiner2011}
map implies a significantly higher reddening of
$E(B-V)~=~0.598~\pm~0.121$.  However, a reddening this high would
increase the photometric temperatures by at least 500 K, which is not
in agreement with the spectroscopic temperatures (see below) or the
temperatures predicted by isochrones.  The predicted stellar
temperatures therefore support the lower reddening from
\citet{Kerber2007}, and indicate that the higher value from
\citet{SchlaflyFinkbeiner2011} is likely due to 100 $\mu$m emission
behind the two target stars.  Bolometric corrections in the $K$ band
were then derived given the photometric temperatures and the empirical
relation from \citet{Buzzoni2010}. NGC~1718 is estimated to be $\sim2$
Gyr old \citep{Kerber2007}; BaSTI isochrones \citep{BaSTIref} with
$[\rm{Fe/H}] = -0.6$ and an age of 2 Gyr indicate that the mass of red
giants in such a GC is about 1.4 M$_{\sun}$.  The photometric
temperatures, bolometric corrections, and turnoff masses then yield
initial photometric surface gravities.

Spectroscopic temperatures and microturbulent velocities ($\xi$, in km
s$^{-1}$) were derived by flattening slopes in \ion{Fe}{1} abundance
with wavelength, reduced equivalent width (REW)\footnote{REW =
  $\log$(EW/$\lambda$), where $\lambda$ is the wavelength of the
  transition.}, and excitation potential (EP, in eV).  As in 
\citet{Fulbright2006}, \citet{KochMcWilliam2008}, and \citet{McW2013},
these [\ion{Fe}{1}/H] abundances are calculated differentially, line
by line, with respect to the [Fe/H] ratios of the cool giant star,
Arcturus.  The large uncertainties in the distance modulus and
reddening make the photometric gravities very uncertain.  For this
reason, scaled solar BaSTI isochrones were utilized for the final
surface gravities, using the spectroscopic temperatures.  The stars
were assumed to be RGB stars; if instead they are AGB stars, the
gravities would change by $<$0.05 dex.  The values from the $z=0.004$,
$[\rm{Fe/H}]~=~-0.66$ and the $z=0.008$, $[\rm{Fe/H}]~=~-0.35$
models were averaged, since the two NGC~1718 stars have
$[\rm{Fe/H}]~\sim~-0.5$.  The final atmospheric parameters are listed
in Table \ref{table:AtmParams}.  Star 9's spectroscopic temperature is
slightly lower than its photometric temperatures, which may be due to
uncertainties in the foreground reddening.

\begin{table*}
\centering
\begin{center}
\caption{Atmospheric parameters.\label{table:AtmParams}}
  \begin{tabular}{@{}lccccc@{}}
  \hline
 & \multicolumn{2}{c}{Photometric} & \multicolumn{2}{c}{Spectroscopic}
  & Isochrone\\
 & $T_{\rm{eff}}$ (K)$^{a}$ & $\log g$ & $T_{\rm{eff}}$ (K) & $\xi$ (km s$^{-1}$) & $\log g$ \\
\hline
NGC 1718-9  & $4049\pm 30$ & $0.70\pm0.2$ & $3820\pm50$ & $1.90\pm0.10$ & $0.52\pm 0.10$\\
NGC 1718-26 & $3858\pm 30$ & $0.80\pm0.2$ & $3890\pm50$ & $1.80\pm0.10$ & $0.67\pm 0.10$ \\
\hline
\end{tabular}
\end{center}
\medskip \raggedright $^{a}$ The errors in the photometric temperature
only consider the \citet{Kerber2007} reddening.
\end{table*}

\begin{figure*}
\begin{center}
\centering
\subfigure[NGC 1718-9]{\includegraphics[scale=0.45]{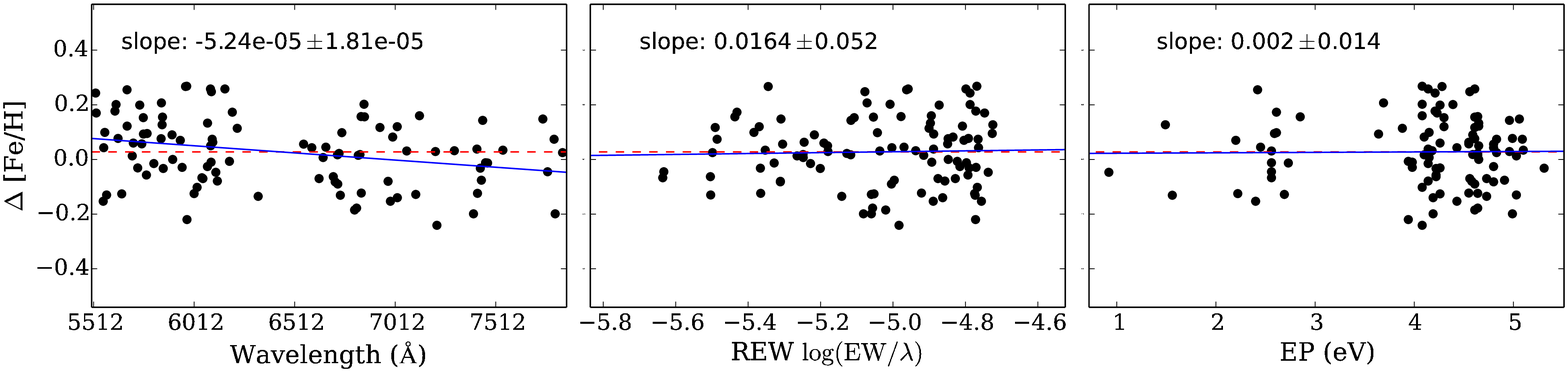}\label{fig:AtmParams9}}
\subfigure[NGC 1718-26]{\includegraphics[scale=0.45]{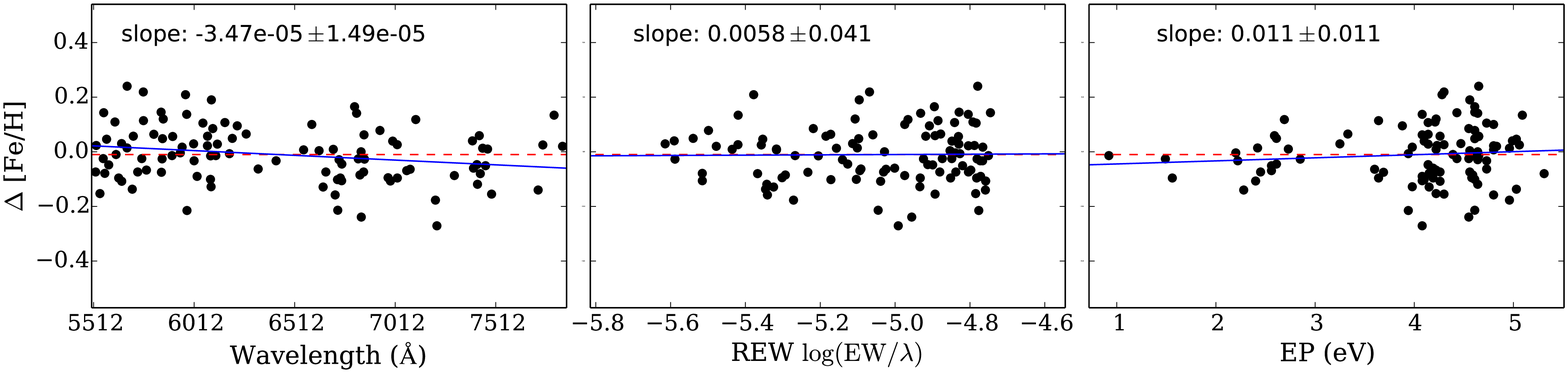}\label{fig:AtmParams26}}
\caption{Trends in $\Delta$[Fe/H] (relative to Arcturus) for NGC
  1718-9 (top) and NGC 1718-26 (bottom).  The solid circles are
  \ion{Fe}{1} lines.  The dashed red line shows the average offset,
  while the solid blue line shows the linear least squares fit.  The
  slopes are quoted in each panel.}\label{fig:AtmParams}
\end{center}
\end{figure*}

\section{Line Lists and Analysis Techniques}\label{sec:LineList}
The line lists of \citet{Fulbright2006,Fulbright2007},
\citet{KochMcWilliam2008}, and \citet{McW2013} were adopted for this
analysis.  The lines in these lists were selected to be clean,
relatively free of blends, and suitable for high precision,
differential analyses.  All Arcturus abundances are calculated
relative to solar abundances derived with the same lines; the EWs from
the above sources were used to derive these abundances.  The LMC
stellar [X/H] abundances from each spectral line are then calculated
relative to the Arcturus [X/H] abundances from the same line. The
average $\Delta [\rm{X/H}]$ offsets of the NGC~1718 stars are then
applied relative to the average [X/H] Arcturus ratios.

All abundances are determined with the July 2014 version of the Local
Thermodynamic Equilibrium (LTE) line analysis code {\tt MOOG}
\citep{Sneden}.  The abundances of Fe, Ca, etc. are determined via
EWs, which are measured with the automated code {\tt DAOSPEC}
\citep{DAOSPECref} for the NGC~1718 stars.  (Recall that the EWs of
\citealt{Fulbright2006,Fulbright2007} and \citealt{KochMcWilliam2008}
were used for the Sun and Arcturus; these EWs were not altered.)  A
moderately-high order polynomial (order 33) was fit to the continuum
levels of the normalized spectra.  Since {\tt DAOSPEC} can have
difficulty accurately measuring the strongest lines (see
\citealt{Sakari2013}), lines with EWs stronger than
100~m\AA \hspace{0.025in} were checked manually with IRAF's {\it
  splot} routine.  For elements with a few lines (e.g., Mg) all EWs
were checked with IRAF's {\it splot} routine.  For most elements,
lines stronger than $\rm{REW}>-4.7$ were not considered because they
are on the (relatively) flat part of the curve of growth and are
therefore sensitive to uncertain damping constants and microturbulent
velocities (see the discussion in \citealt{McWilliam1995}).  For
elements with only strong lines and HFS components, this limit was
pushed to $\rm{REW}=-4.5$.  Although this may introduce uncertainties
$\sim 0.1$ dex, tests were performed to ensure that the lines were not
saturated.  Random errors in EW-based abundances were determined as in
\citet{Shetrone2003}.  The lines utilized for EWs are shown in Table
\ref{table:EWs}, along with the values measured in NGC 1718-9 and
-26.

\begin{table}
\centering
\begin{center}
\caption{EW Line list.\label{table:EWs}}
  \newcolumntype{d}[1]{D{,}{\pm}{#1}}
  \begin{tabular}{@{}ccccc@{}}
  \hline
Wavelength & Element & EP  & \multicolumn{2}{c}{EW (m\AA)}\\
     (\AA) &         & (eV)& NGC 1718-9 & NGC 1718-26\\
\hline
5522.450 & 26.0 & 4.210  & 90.10  &  72.70 \\
5525.540 & 26.0 & 4.230  & 98.90  &  90.00 \\
5543.940 & 26.0 & 4.220  & -      &  91.00 \\
5560.210 & 26.0 & 4.430  & 71.80  &  78.80 \\
5562.710 & 26.0 & 4.430  & 95.80  & 100.00 \\
\hline
\end{tabular}
\end{center}
\medskip
\raggedright {\bf Notes: } Table \ref{table:EWs} is published in
its entirety in the electronic edition of \textit{Monthly Notices of
  the Royal Astronomical Society}. A portion is shown here for
guidance regarding its form and content.\\
{\bf References:} Lines were selected from the
line lists of \citet{Fulbright2006,Fulbright2007} and
\citet{KochMcWilliam2008}; the solar and Arcturus EWs are also taken
from those papers.\\
\end{table}

\begin{figure*}
\begin{center}
\centering
\includegraphics[scale=0.7]{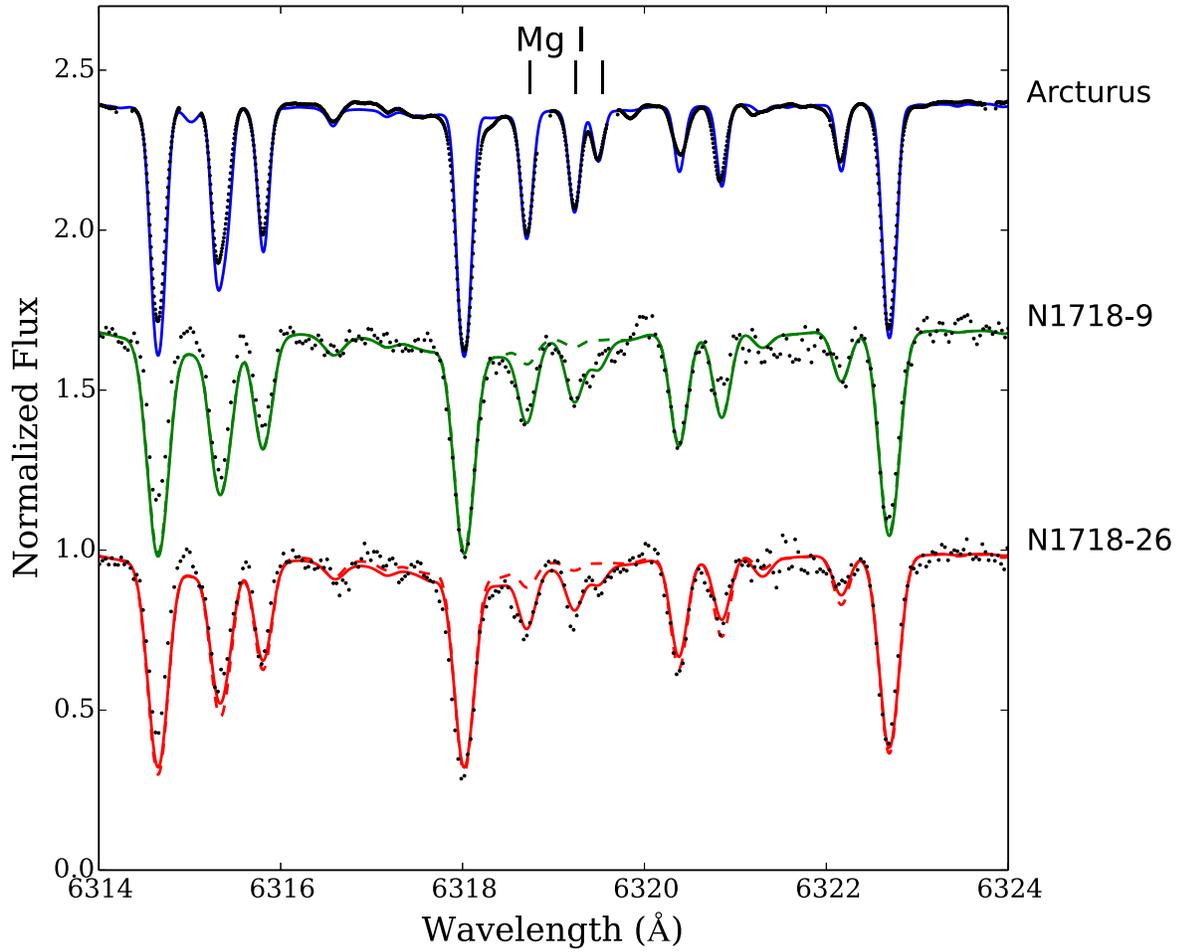}
\caption{Syntheses of the 6318/6319 \AA \hspace{0.025in} \ion{Mg}{1}
  lines in Arcturus (top), NGC~1718-9 (middle), and NGC~1718-26
  (bottom).  The solid lines show the best-fit abundances to the 6318
  \AA \hspace{0.025in} line; dashed lines show NGC~1718's integrated
  light value from \citet{Colucci2012}.}\label{fig:Mg}
\end{center}
\end{figure*}

Other abundances are derived via spectrum syntheses (SS); in this
case, spectral lines in a given wavelength range were selected from the Kurucz
database.\footnote{\url{http://kurucz.harvard.edu/linelists.html}}
Molecular lines were included in regions where they are noted in the
Arcturus atlas \citep{Hinkle2003}.  Syntheses of the sun and Arcturus
were first performed to identify the solar and Arcturus
synthesis-based abundances.   The uncertainty in
SS-based abundances is determined from the range of abundances that
can fit a given spectral line profile.  The lines used for SS are
shown in Table \ref{table:SSLinelist}, along with the abundances for
the sun, Arcturus, and the NGC 1718 stars.  Figure \ref{fig:Mg} shows
examples of the fits to the 6318/6319 \AA \hspace{0.025in} \ion{Mg}{1}
lines, which are near a \ion{Ca}{1} autoionization feature.

Appendix \ref{appendix:Errors} presents a detailed analysis of the
abundance sensitivity to the adopted atmospheric parameters, following
the procedure of \citet{McWilliam1995,McW2013}.  The final errors in
[X/Fe] ratios are shown in Table \ref{table:finalerrors}.

\begin{table*}
\centering
\begin{minipage}{165mm}
\begin{center}
\caption{SS Line list.\label{table:SSLinelist}}
  \newcolumntype{d}[1]{D{,}{\pm}{#1}}
  \begin{tabular}{@{}cccccd{4}d{6}@{}}
  \hline
Wavelength & Element & EP  & \multicolumn{1}{l}{$\log \epsilon$ (X)} & & &\\
     (\AA) &         & (eV)& Sun & Arcturus & \multicolumn{1}{c}{NGC 1718-9} & \multicolumn{1}{c}{NGC 1718-26}\\
\hline
6300.304 &  8.0 &  0.00 & 8.84 & 8.69 & 8.24,0.07 & 8.29,0.05\\
6154.222 & 11.0 &  2.10 & 6.28 & 5.87 & 5.54,0.10 & 5.64,0.10\\
6160.746 & 11.0 &  2.10 & 6.33 & 5.94 & 5.64,0.10 & 5.44,0.10\\
6318.705 & 12.0 &  5.10 & 7.60 & 7.43 & 7.15,0.10 & 7.10,0.10\\
6319.232 & 12.0 &  5.10 & 7.60 & 7.45 & 7.25,0.10 & 7.20,0.10\\
6696.015 & 13.0 &  3.14 & 6.30 & 6.20 & 5.68,0.10 & 5.75,0.05\\
6698.665 & 13.0 &  3.14 & 6.32 & 6.13 & 5.65,0.10 & 5.75,0.10\\
5782.110$^{a}$ & 29.0 &  1.64 & 4.22 & 3.94 & 3.04,0.10 & 3.14,0.10\\
6645.130 & 63.1 &  1.37 & 0.42 & 0.24 & 0.13,0.05 & 0.19,0.05 \\
\hline
\end{tabular}
\end{center}
\end{minipage}\\
\medskip
\raggedright $^{a}$ HFS components were included in the syntheses. \\
\end{table*}

\section{Abundances}\label{sec:Abunds}
Final abundances are shown in Table \ref{table:Abunds}.
As discussed in Sections \ref{sec:AtmParams} and \ref{sec:LineList},
all abundances are calculated differentially with respect to Arcturus
and the Sun.  All [X/Fe] ratios are calculated using \ion{Fe}{1}.  Many analyses
utilize \ion{Fe}{2} for singly ionized species (and \ion{O}{1}),
because the systematic errors are expected to be similar.  However,
Table \ref{table:finalerrors} demonstrates that for the NGC~1718
stars, the uncertainties are smaller when the [X/Fe] ratios are
calculated with \ion{Fe}{1} (see Appendix \ref{appendix:Errors}).

\begin{table*}
\centering
\begin{minipage}{165mm}
\begin{center}
\caption{Derived abundances and random errors; total errors are given in Table \ref{table:finalerrors}.\label{table:Abunds}}
  \newcolumntype{d}[1]{D{,}{\pm}{#1}}
  \begin{tabular}{@{}ld{3}cccd{3}cccd{3}cc@{}}
  \hline
 & \multicolumn{3}{c}{Arcturus} & & \multicolumn{3}{c}{NGC~1718-9} & & \multicolumn{3}{c}{NGC~1718-26}\\
 & \multicolumn{1}{c}{[X/Fe]} & $N$ & Method & & \multicolumn{1}{c}{[X/Fe]} & $N$ & Method & & \multicolumn{1}{c}{[X/Fe]} & $N$ & Method \\
\hline
\ion{Fe}{1} & -0.53,0.02 & 152 & EW    & & -0.55,0.01 &  99 & EW & & -0.54,0.01 & 103 & EW\\
\ion{Fe}{2} & -0.45,0.03 &  5  & EW    & & -0.54,0.01 &   2 & EW & & -0.57,0.03 &   2 & EW\\
\ion{O}{1}  &  0.30,0.05 &  1  & SS    & & -0.13,0.07 &   1 & SS & & -0.11,0.05 &   1 & SS\\
\ion{Na}{1} &  0.13,0.03 &  2  & SS    & & -0.13,0.07 &   2 & SS    & &-0.18,0.09 &   2 & SS\\
\ion{Mg}{1} &  0.36,0.06 &  11 & EW/SS & &  0.11,0.04 &   7 & EW/SS & & 0.11,0.03 &   7 & EW/SS\\
\ion{Al}{1} &  0.41,0.05 &  5  & EW/SS & &  0.01,0.07 &   4 & EW/SS & & 0.04,0.03 &   4 & EW/SS\\
\ion{Si}{1} &  0.30,0.02 & 19  & EW    & &  0.11,0.03 &   9 & EW & &  0.13,0.04 &  12 & EW\\
\ion{Ca}{1} &  0.20,0.02 & 14  & EW    & &  0.09,0.10 &   2 & EW & &  0.11,0.07 &   2 & EW\\
\ion{Ti}{1} &  0.27,0.02 & 25  & EW    & &  0.09,0.03 &   7 & EW & &  0.06,0.03 &  12 & EW\\
\ion{Ti}{2} &  0.20,0.02 &  6  & EW    & & -0.10,0.10 &   2 & EW & & -0.06,0.02 &   2 & EW\\
\ion{V}{1}  &  0.09,0.03 &  2  & EW    & & -0.09,0.08 &   3$^{a}$ & EW & & -0.06,0.04 & 3$^{a}$ & EW\\
\ion{Mn}{1} & -0.12,0.04 &  5  & EW    & & -0.19,0.12 &   3$^{a}$ & EW & & -0.22,0.08 &   3$^{a}$ & EW\\
\ion{Ni}{1} &  0.11,0.02 & 17  & EW    & & -0.02,0.05 &  15 & EW & & -0.02,0.05 &  14 & EW\\
\ion{Cu}{1} &  0.25,0.10 &  1  & SS    & & -0.63,0.10 &   1 & SS & & -0.49,0.10 &   1 & SS\\
\ion{Rb}{1} &  0.03,0.02 &  2  & EW    & & -0.24,0.09 &   2 & EW & & -0.25,0.13 &   2 & EW\\
\ion{Y}{2}  & -0.09,0.07 &  3  & EW    & & -0.04,0.08 &   2 & EW & & -0.06,0.08 &   1 & EW\\
\ion{Zr}{1} & -0.25,0.04 &  4  & EW    & & -0.18,0.06 & 3$^{a}$ & EW & & -0.05,0.06 & 3$^{a}$ & EW\\
\ion{La}{2} & -0.05,0.04 &  5  & EW    & &  0.27,0.07 &   4 & EW & &  0.30,0.10 &   3 & EW\\
\ion{Eu}{2} &  0.27,0.05 &  1  & SS    & &  0.22,0.05 &   1 & SS & &  0.26,0.05 &   1 & SS\\
\hline
\end{tabular}
\end{center}
\end{minipage}\\
\medskip
\raggedright {\bf Notes: } $^{a}$ Lines have $-4.7<$REW$<-4.5$; in all
cases these lines have HFS components.\\
\end{table*}

\subsection{Iron}\label{subsec:FeAbunds}
The \ion{Fe}{1} abundances in the two NGC~1718 stars are derived from
EWs of $\sim100$ unblended lines, while only two \ion{Fe}{2} lines are
measurable.  The greater number of \ion{Fe}{1} lines means that
\ion{Fe}{1} has a lower random error than \ion{Fe}{2}; still,
\ion{Fe}{1} and \ion{Fe}{2} are in good agreement in all cases.  In
particular, any offsets that may be expected from NLTE effects (e.g.,
\citealt{KraftIvans2003}) are minimized with this differential
abundance approach (since Arcturus is expected to have the same NLTE
corrections as the NGC~1718 stars).

\begin{figure*}
\begin{center}
\centering
\includegraphics[scale=0.8]{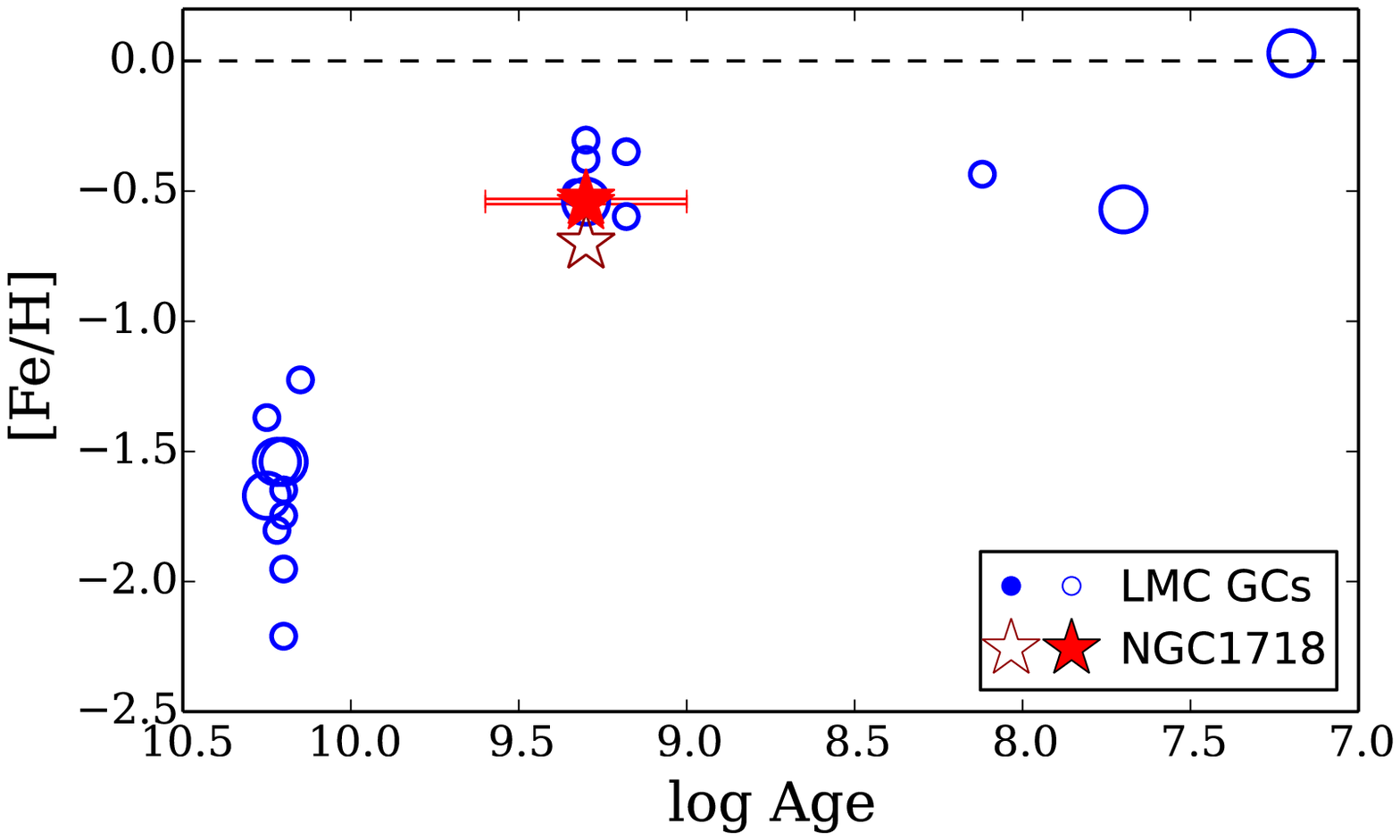}
\caption{The age-metallicity relation for LMC GCs.  The ages are
  from \citet{Baumgardt2013}.  The abundances of the two individual
  NGC~1718 stars are shown as solid red stars, while NGC~1718's IL
  abundance from \citet{Colucci2012} is shown as an open star.
  Abundances of other LMC GCs are also shown.  The individual stars
  analyzed by
  \citet{Mucciarelli2008,Mucciarelli2010,Mucciarelli2011,Mucciarelli2014}
  and \citet{Johnson2006} are averaged together for each cluster and
  are shown as small, open circles.  The IL values for other clusters
  from \citet{Colucci2012} are shown as large open
  circles.}\label{fig:AMR}
\end{center}
\end{figure*}

The abundances from the two NGC~1718 stars indicate an average cluster
metallicity of $[\rm{Fe\;I/H}] = -0.55\pm 0.01$.  This value roughly
agrees with the isochrone fits by \citet{Kerber2007}, and is
consistent with the age/metallicity relations of other LMC GCs (see
Figure \ref{fig:AMR} and \citealt{MackeyGilmore2003}) and field
stars (e.g., \citealt{Piatti2012}).  However, this value is higher
than the integrated light metallicity from
\citet{Colucci2011,Colucci2012}, who find
$[\rm{Fe\;I/H}]~=~-~0.70~\pm~0.05$ (though note that their
$[\rm{Fe\;II/H}]~=~-0.26\pm 0.18$) and the calcium triplet
measurements from \citet{Grocholski2006}, who find
$[\rm{Fe/H}]~=~-0.80\pm0.03$. Integrated abundances can be extremely
sensitive to the properties of the adopted isochrone (e.g.,
\citealt{Sakari2014,Colucci2016}). In fact, for NGC~1718, Colucci et
al. find two appropriate isochrones that reproduce their \ion{Fe}{1}
line strengths: one with an age of 1 Gyr and $[\rm{Fe/H}] = -0.39$,
the other with age = 2.5 Gyr and $[\rm{Fe/H}] = -0.89$.  Because their
analysis was unable to distinguish between the solutions, Colucci et
al. averaged the age and metallicity, deriving a final abundance of
$[\rm{Fe\;I/H}]~=~-0.70~\pm~0.05$.  Uncertainties in the age and
metallicity may introduce uncertainties in the integrated
[\ion{Fe}{1}/H] ratio $\ga 0.1$ dex
\citep{Sakari2014,Colucci2016}.  The disagreement with the CaT
metallicities could be due to NGC~1718's lower [$\alpha$/Fe], such
that the CaT indicates a low [Z/H] rather than [Fe/H].

\subsection{Light Elements: O, Na, Mg, and Al}\label{subsec:LightAbunds}
The O and Na abundances are derived solely with SSs. The forbidden
line at 6300 \AA \hspace{0.025in} is used to determine the \ion{O}{1}
abundances.  CN lines are included in these syntheses, and the O
abundances are therefore mildly sensitive to the adopted C abundance.
The Na abundances are from the 6154/6160 \AA \hspace{0.025in} doublet,
which should be least sensitive to non-LTE effects in this metallicity
range \citep{Lind2011}; the INSPECT database\footnote{Data obtained
  from the INSPECT database, version 1.0:
  \url{http://www.inspect-stars.com/}} confirms that any corrections
should be small.  At most, non-LTE effects would introduce offsets
$\sim 0.1$ dex in these giant stars \citep{Mashonkina2000}.

Mg and Al are derived with EWs and SSs.  Most of the Mg
and Al lines are sufficiently clean for EW analyses, with two
exceptions: the 6318 and 6319 \AA \hspace{0.025in} \ion{Mg}{1} lines
and the 6696 and 6698 \AA \hspace{0.025in} \ion{Al}{1} lines.  In the
latter case the lines are weak enough to make EW measurements
difficult.  The 6318 and 6319~\AA \hspace{0.025in} \ion{Mg}{1} lines
are fairly strong---however, they are located on top of a broad
\ion{Ca}{1} autoionization feature, which makes continuum
identification difficult.  Figure \ref{fig:Mg} shows these syntheses
in Arcturus and the two NGC~1718 stars. 

The only previous detailed abundances for NGC~1718 are from the
integrated light analysis of \citet{Colucci2012}. The integrated
[Na/Fe] and [Al/Fe] ratios are slightly higher than the individual
stars, but are in agreement within the errors.  The integrated
$[\rm{Mg/Fe}] = -0.90\pm0.30$ is considerably lower than the
individual abundances. In their analysis of MW GCs,
\citet{Colucci2016} find that their IL [Mg/Fe] ratios are generally
$\sim0.2-0.3$ dex lower than the values from individual values.  This
offset may be due to systematic effects (e.g., NLTE effects, as
proposed by Colucci et al.) or it may reflect real abundance
variations within the cluster.  This possibility will be discussed in
more detail in Section \ref{subsec:AbundSpreads}.

\subsection{$\alpha$-Elements}\label{subsec:AlphaAbunds}
The Si, Ca, and Ti abundances are determined from EWs.  In the
NGC~1718 stars there are 8-12 \ion{Si}{1} lines, but only two
\ion{Ca}{1} lines (there are many \ion{Ca}{1} lines in this wavelength
range, but most are too strong for this analysis; see Section
\ref{sec:LineList}).  There are 7-12 \ion{Ti}{1} lines available in
the NGC~1718 spectra, but only 2 \ion{Ti}{2} lines.  In NGC~1718-9,
the \ion{Ti}{2} abundance is lower than \ion{Ti}{1}. Si, Ca, and
\ion{Ti}{1} are all in excellent agreement, suggesting that NGC~1718
is a moderately $\alpha$-enhanced cluster.

From the integrated light spectrum, \citet{Colucci2012} derive
$[\rm{Ca/Fe}] = -0.14\pm0.14$, which is lower than the individual
stars analyzed in this work.  Colucci et al.'s $[\rm{Ti\; I}/Fe]=0.7$
is much higher than NGC~1718-9 and -26; this high abundance suggests
that the integrated Ti is systematically offset in some way.
The optical IL \ion{Ti}{1} lines are very sensitive to stochastic
sampling effects \citep{Sakari2014}; similarly, \citet{Colucci2016}
find a large scatter in the IL \ion{Ti}{1} abundances of their MW
GCs.

\subsection{Iron-peak Elements}\label{subsec:FePeakAbunds}
EWs are utilized for abundances of V, Mn, and Ni, though [Mn/Fe] was
also verified with SSs.  Cu was determined solely with SSs.  V, Mn,
and Cu require HFS components to properly account for the strengths of
the lines. The HFS line lists and the Arcturus EW measurements from
\citet{McW2013} are adopted here.  Note that two of the three
\ion{V}{1} lines and all three of the \ion{Mn}{1} lines have REWs
larger than $-4.7$.  However, these lines all have HFS splitting which
de-saturates the lines; the EWs of such features are therefore still
sensitive to the abundance. Tests of these features show that the
lines are formed throughout the atmospheres and are not saturated in
the top few layers, where the models are least reliable.  The Cu
abundance is determined from the 5782 \AA \hspace{0.025in} line.  HFS
components and isotopic splits were included in the syntheses---again,
these lists are from \citet{McW2013}.  A solar isotopic ratio of
$^{63}$Cu/$^{65}$Cu$=2.24$ is adopted \citep{Asplund2009}. In this
metallicity range, non-LTE effects are not expected to be significant
for these Cu lines \citep{Yan2015}.

The abundances of the two stars indicate that NGC~1718 has solar
[Ni/Fe], mildly subsolar [V/Fe] and [Mn/Fe], and very deficient
[Cu/Fe].  Colucci et al.'s integrated light analysis suggested that
the abundances of iron-peak elements Mn and Ni are roughly solar in
NGC~1718.  Although the individual stars show a lower [Mn/Fe], the
results are in agreement within the errors.

\subsection{Neutron-Capture Elements}\label{subsec:NeutronCaptureAbunds}
Neutron-capture elements form when free neutrons are captured by seed
nuclei.  The build-up of neutrons in the nucleus leads to heavier
isotopes, while the subsequent decay into protons gradually forces
elements to higher atomic number.  The types of elements and isotopes
that form from neutron captures depend on the incoming neutron flux,
i.e., the slow (s-) neutron-capture process has different
nucleosynthetic yields than the rapid (r-) process. Although the heavy
elements typically form in both processes, certain elements form
primarily in only one of the processes.

\subsubsection{s-Process Elements}\label{subsec:sAbunds}
In the sun, Y, Zr, and La form primarily in the s-~process, while
$\sim50$\% of Rb forms via the s-process
\citep{Burris2000}. Abundances of all four elements were determined
from EWs, utilizing the HFS line lists from \citet{McW2013}.  Two
\ion{Rb}{1} lines are utilized, at 7800 and 7947 \AA; these lines are
of moderate strength in both stars.  Two \ion{Y}{2} lines (5728 and
7450 \AA) are detectable in NGC~1718-9, while only the bluer one is
detectable in NGC~1718-26.  Zr abundances are determined from three
strong lines, all with $-4.7<\rm{REW}<-4.5$; again, these lines have
HFS components, and tests were done to ensure that the individual
components were not saturated.  Three to four moderate strength
\ion{La}{2} lines were utilized.

\citet{Colucci2012} determine an integrated [Y/Fe] for NGC~1718, which
they find to be roughly solar.  This agrees with the abundances from
the two individual stars in this work.  Though they do not detect La
lines in NGC~1718, Colucci et al. do present an integrated Ba
abundance of $[\rm{Ba/Fe}]~=~0.20\pm 0.30$.  Barium is also primarily
an s-process element in this metallicity range, and the lines are
easily detectable; unfortunately, in this analysis the barium lines
are far too strong in both of the target stars for a reliable
abundance analysis (see Section \ref{sec:LineList}). Given that they
are both second peak, s-process elements, Ba is expected to track La.
Though this analysis finds slightly higher [La/Fe], the two are in
agreement within the errors.

\subsubsection{r-Process Elements}\label{subsec:rAbunds}
Eu is primarily an r-process element, with only 3\% forming from the
s-process in the sun \citep{Burris2000}.  Eu was determined
from spectrum syntheses of the 6645~\AA \hspace{0.025in} line. Nearby
CN features were included in the syntheses.  The HFS and isotopic
information from \citet{McW2013} were included, and the solar
$^{153}\rm{Eu}/^{151}\rm{Eu} = 1.09$ ratio was assumed
\citep{Asplund2009}.  Note that \citet{Colucci2012} do not provide a
Eu abundance for NGC~1718.  In this work, NGC~1718 is found to show
moderate r-process enhancement.

\subsection{Lithium}\label{subsec:Li}
\begin{figure*}
\begin{center}
\centering
\includegraphics[scale=0.6,trim=0 0.5in 0 0.5in,clip]{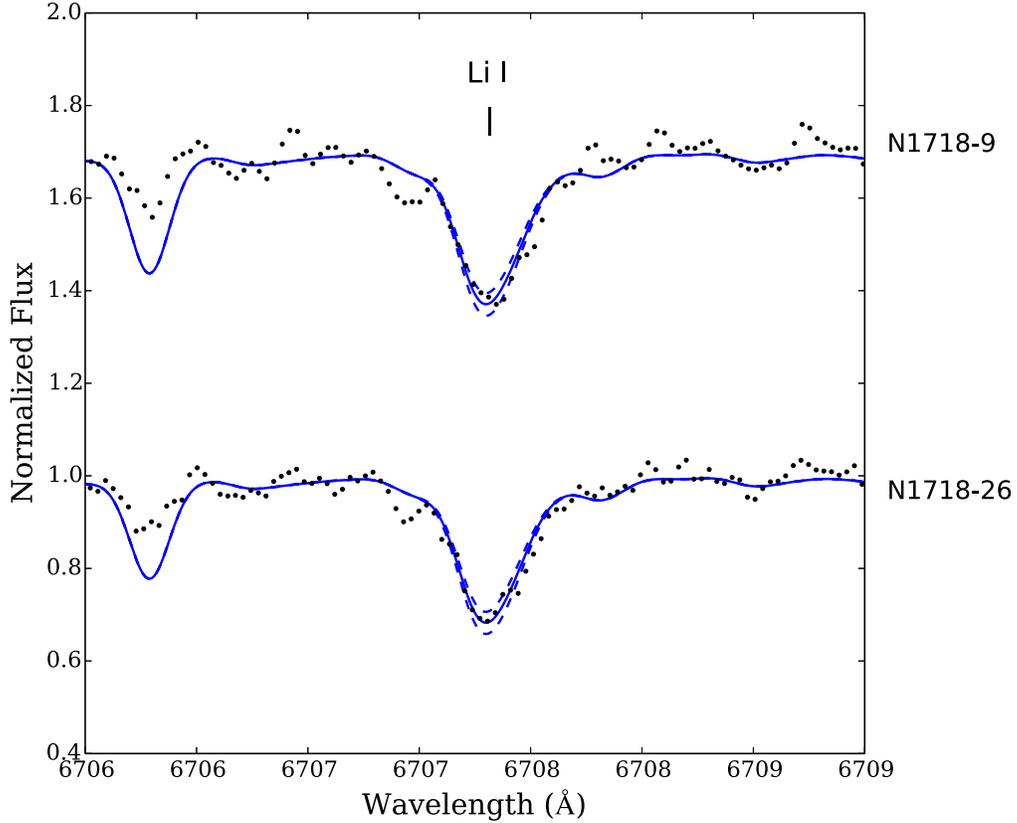}
\caption{Syntheses of the 6707 \AA \hspace{0.025in} \ion{Li}{1} line
  in the NGC~1718 stars.  The solid lines show the best-fit abundances,
while the dashed lines show the $\pm 1\sigma$ uncertainties.}\label{fig:Li}
\end{center}
\end{figure*}

Lithium is detectable in the two NGC~1718 stars, as shown in Figure
\ref{fig:Li}. Syntheses were done utilizing lines and HFS components
from the Kurucz
database\footnote{\url{http://kurucz.harvard.edu/linelists/}} and
considering only $^{7}$Li lines.  The derived abundances are
$\log\epsilon(\rm{Li})_{\rm{LTE}}~=~0.05\pm0.05$ and
$\log\epsilon(\rm{Li})_{\rm{LTE}}~=~0.15\pm0.05$ for NGC~1718~-~9 and NGC~1718~-~26,
respectively.  Note that NLTE corrections are expected to be $\sim +0.4$
dex for these stars, based on estimates from the INSPECT
database\footnote{Data obtained from the INSPECT database, version 1.0:
  \url{http://www.inspect-stars.com/}} \citep[though
  note that the atmospheric parameters in the database do not extend
  to sufficiently cool temperatures and low surface
  gravities]{Lind2009}.  These rough Li
abundances seem appropriate for normal, evolved RGB stars in GCs
(e.g., \citealt{Mucciarelli2014Li,D'Orazi2014,D'Orazi2015,Kirby2016}).

\section{Discussion}\label{sec:Discussion}
As discussed in Section \ref{sec:Intro}, NGC~1718 is a valuable target
for chemical abundance analyses, for two reasons: 1) its status as an
intermediate-age, sparse GC (which will be discussed in Section
\ref{subsec:AbundSpreads}), and 2) its presence in the LMC, which can
be used to probe the chemical evolution of the LMC (see Section
\ref{subsec:LMC}).

\subsection{NGC 1718 as a Globular Cluster}\label{subsec:AbundSpreads}
Section \ref{sec:Intro} described the presence of multiple populations
in GCs.  To summarize, in the Milky Way, all classical GCs show
star-to-star  abundance variations in Na and O; some also show
variations in Mg and Al (e.g., \citealt{Carretta2009}).  Though Mg and
Al variations are typically only seen in massive, metal-poor GCs
(e.g., M15; \citealt{Sneden1997}), recent observations in the $H$-band
suggest that the Mg/Al anticorrelation may be present in metal-rich
GCs as well \citep{Meszaros2015}.  No convincing signs of similar
multiple populations have yet been detected in intermediate-age or
young LMC GCs (e.g.,
\citealt{Mucciarelli2008,Mucciarelli2011,Mucciarelli2014}) though they
have been detected in old LMC GCs
\citep{Johnson2006,Mucciarelli2009}. Abundance variations within
distant GCs can be {\it inferred} from integrated spectra, e.g.,
through high integrated [Na/Fe] ratios
\citep{Sakari2013,Sakari2015,Colucci2014}. \citet{Colucci2012} find
that the intermediate-age GCs do not have high integrated [Na/Fe]
ratios, while the older GCs do.  Though they did find low integrated
[Mg/Fe] in NGC~1718, they attributed this to low primordial Mg
rather than to star-to-star variations within the cluster.

\begin{figure*}
\begin{center}
\centering
\subfigure{\includegraphics[scale=0.6,trim=1.0in 0in 0.05in 0.0in]{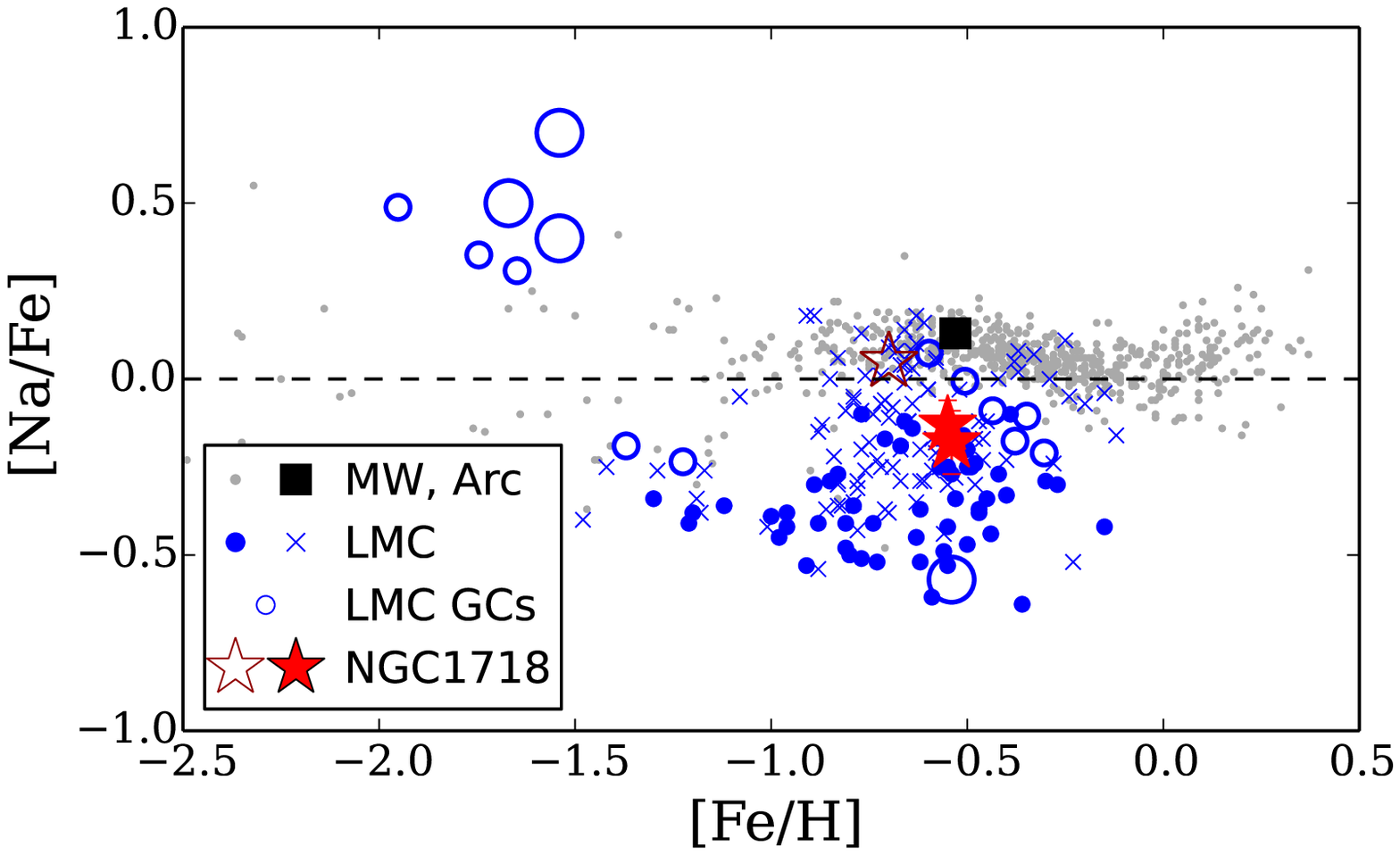}\label{fig:NaFe}}
\subfigure{\includegraphics[scale=0.6,trim=0.7in 0in 1.25in 0.0in]{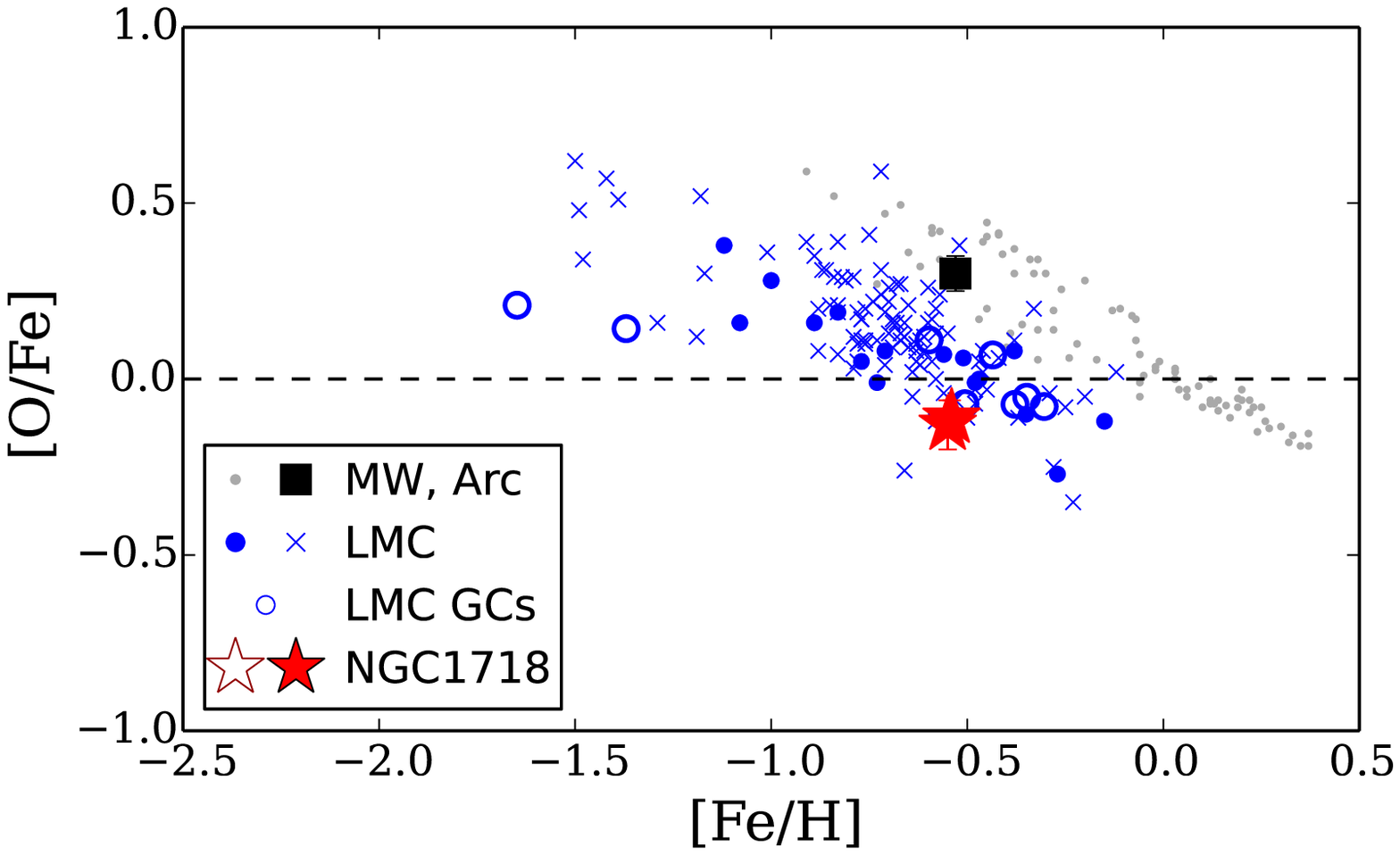}\label{fig:OFe}}
\caption{[Na/Fe] (left) and [O/Fe] (right) ratios in NGC~1718 compared
  to MW and LMC field stars and other LMC GCs.  Red stars show the two
  NGC~1718 stars from this analysis, along with the random errors.
  The large maroon open star shows NGC~1718's integrated light
  abundance from \citet{Colucci2012}.  The large black square shows
  the Arcturus value derived in this paper.  Grey points are MW field
  stars from \citet{Venn2004}, with supplements from
  \citet{Reddy2006} and \citet{Bensby2005}---note that the O
  abundances from Reddy et al. are not shown, because they require
  NLTE corrections.  Small blue crosses are the LMC bar stars from
  \citet{VanderSwaelmen2013}, while small filled blue circles are the
  disk stars from \citet{Pompeia2008} which have been reanalyzed by
  \citet{VanderSwaelmen2013}.  LMC GCs are shown as open circles:
  large circles are integrated light values from \citet{Colucci2012}
  while small circles are averages of individual stars from
  \citet{Johnson2006} and
  \citet{Mucciarelli2008,Mucciarelli2010,Mucciarelli2011,Mucciarelli2014}.
}\label{fig:NaOfigs}
\end{center}
\end{figure*}

\begin{figure*}
\begin{center}
\centering
\subfigure{\includegraphics[scale=0.6,trim=1.0in 0in 0.05in 0.0in]{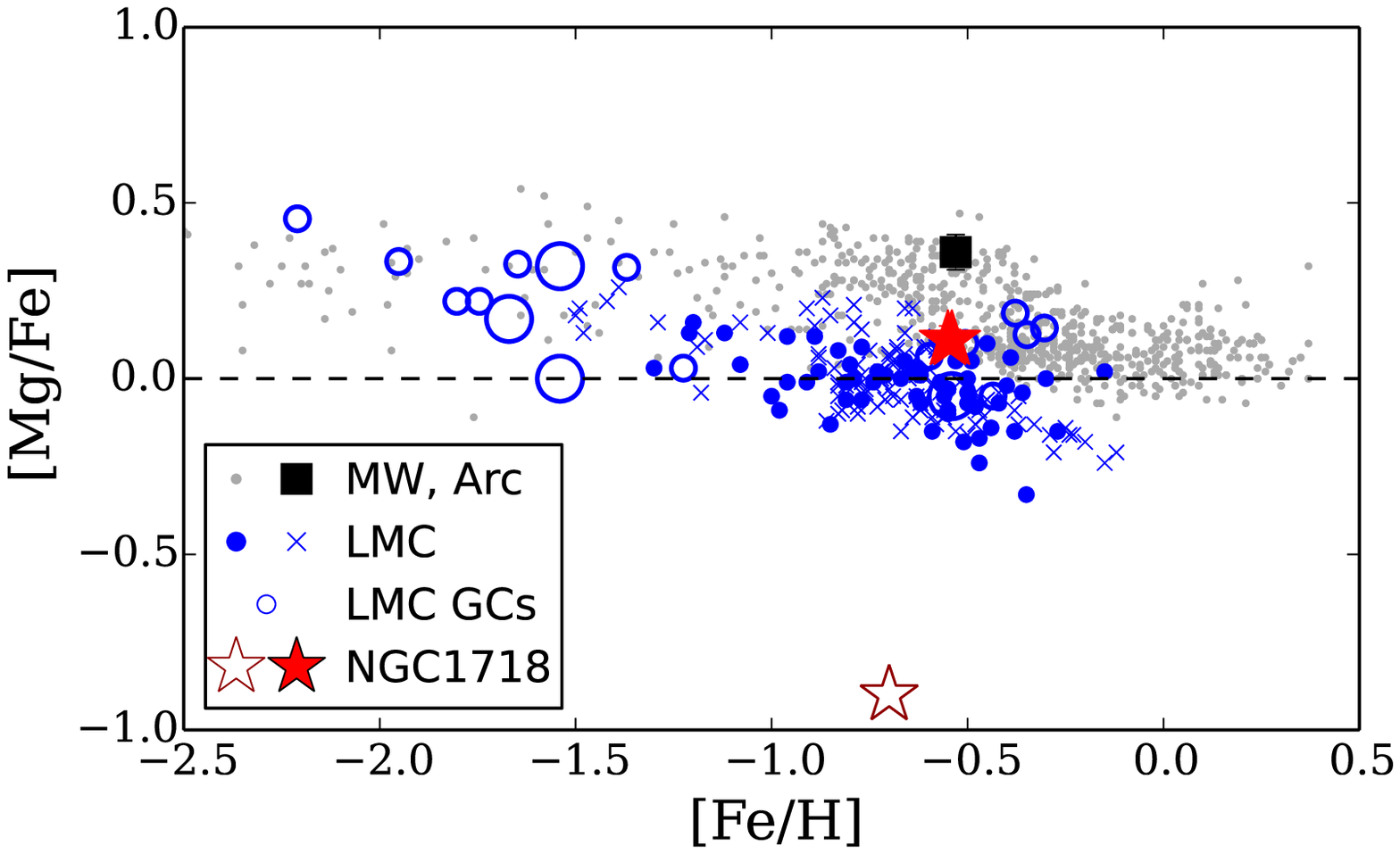}\label{fig:MgFe}}
\subfigure{\includegraphics[scale=0.6,trim=0.7in 0in 1.25in 0.0in]{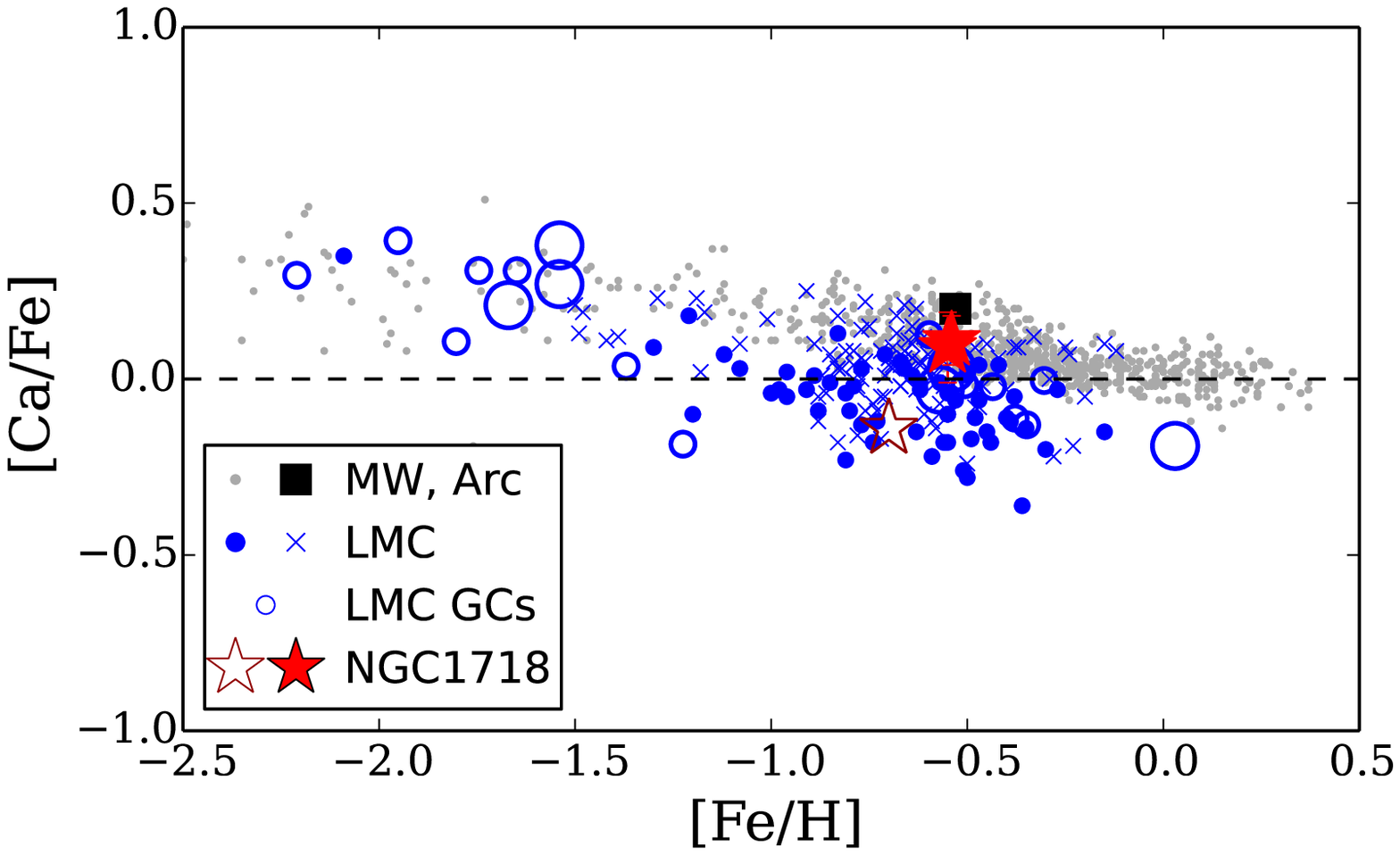}\label{fig:CaFe}}
\caption{[Mg/Fe] (left) and [Ca/Fe] (right) ratios in NGC~1718
  compared to MW and LMC field stars and other LMC GCs.  Points are as
  in Figure \ref{fig:NaOfigs}.  The left panel shows that the
  previously-derived low [Mg/Fe] in NGC~1718 from IL spectroscopy is
  not supported by the two individual stars in this
  analysis.}\label{fig:MgAlfigs}
\end{center}
\end{figure*}

With only two stars in this sample it is difficult to assess the
presence of any abundance spreads within NGC~1718.  Though the two
target stars do appear to have slightly different O and Na abundances
(with one more O-deficient and Na-enhanced than the other) the
abundances are identical within random errors.  The Na abundance
differences are driven solely by the 6160 \AA \hspace{0.025in} line
(the 6154 \AA \hspace{0.025in} line gives similar Na abundances).
Thus, there is no convincing evidence for a spread in Na and O between
the two stars.  Additionally, neither of the stars have enhanced Na or
deficient O relative to the LMC field stars (see Figure
\ref{fig:NaOfigs}).  Similarly, the Mg and Al abundances in the two
NGC~1718 stars track the LMC field star distribution and are identical
within the errors (see Figure \ref{fig:MgFe}).

The O, Na, Mg, and Al abundances of these stars therefore appear to
follow the ``primordial'' abundance signature of the cloud from which
NGC~1718 formed.  This lack of significant abundance differences
between these two stars does not rule out the presence of multiple
populations in NGC~1718---indeed, Colucci et al.'s higher [Na/Fe] and
[Al/Fe] and lower [Mg/Fe] integrated ratios suggest that observations
of more cluster stars are necessary to resolve this issue.  It does
imply, however, that Colucci et al.'s unexpected integrated abundances
are not likely to be the result of unusual chemical evolution
in the LMC.

\subsection{The Chemical Evolution of NGC~1718, the LMC, and its GC System}\label{subsec:LMC}
The agreement between the abundances of the NGC~1718 stars and the LMC
field stars suggests that NGC~1718 is a valuable probe for examining
the chemical evolution of the LMC, particularly for elements like Mn
and Rb, which have not yet been extensively studied in the LMC GCs.
NGC~1718 can be compared with other stars and GCs, both in the LMC and
the MW; however, systematic uncertainties may complicate these
comparisons.  The differential nature of this analysis, however, means
that comparisons with Arcturus will be extremely robust.

\subsubsection{$\alpha$-elements}\label{subsubsec:DiscussAlpha}
One major result from this analysis is that NGC~1718's abundances do
not require a unique formation scenario from the pure ejecta of Type
Ia supernovae, as proposed by \citet{Colucci2012}.  The normal [Mg/Fe]
in the two stars (relative to the LMC field stars) suggests that the
low integrated [Mg/Fe] is either 1) an indication of a severe Mg
spread within NGC~1718 (see Section \ref{subsec:AbundSpreads}) or 2) a
result of some systematic uncertainty in the IL analysis.  Regardless
of the cause for the discrepancy with the integrated abundance, the
normal [Mg/Fe] in the two stars from this analysis indicates that
NGC~1718's primordial abundances were not distinct from the LMC field
stars at the same metallicity.  In the other $\alpha$-elements, e.g.,
Ca (see Figure \ref{fig:CaFe}) NGC~1718 also tracks the LMC field
stars.

\begin{figure*}
\begin{center}
\centering
\subfigure{\includegraphics[scale=0.6,trim=1.0in 0in 0.05in 0.0in]{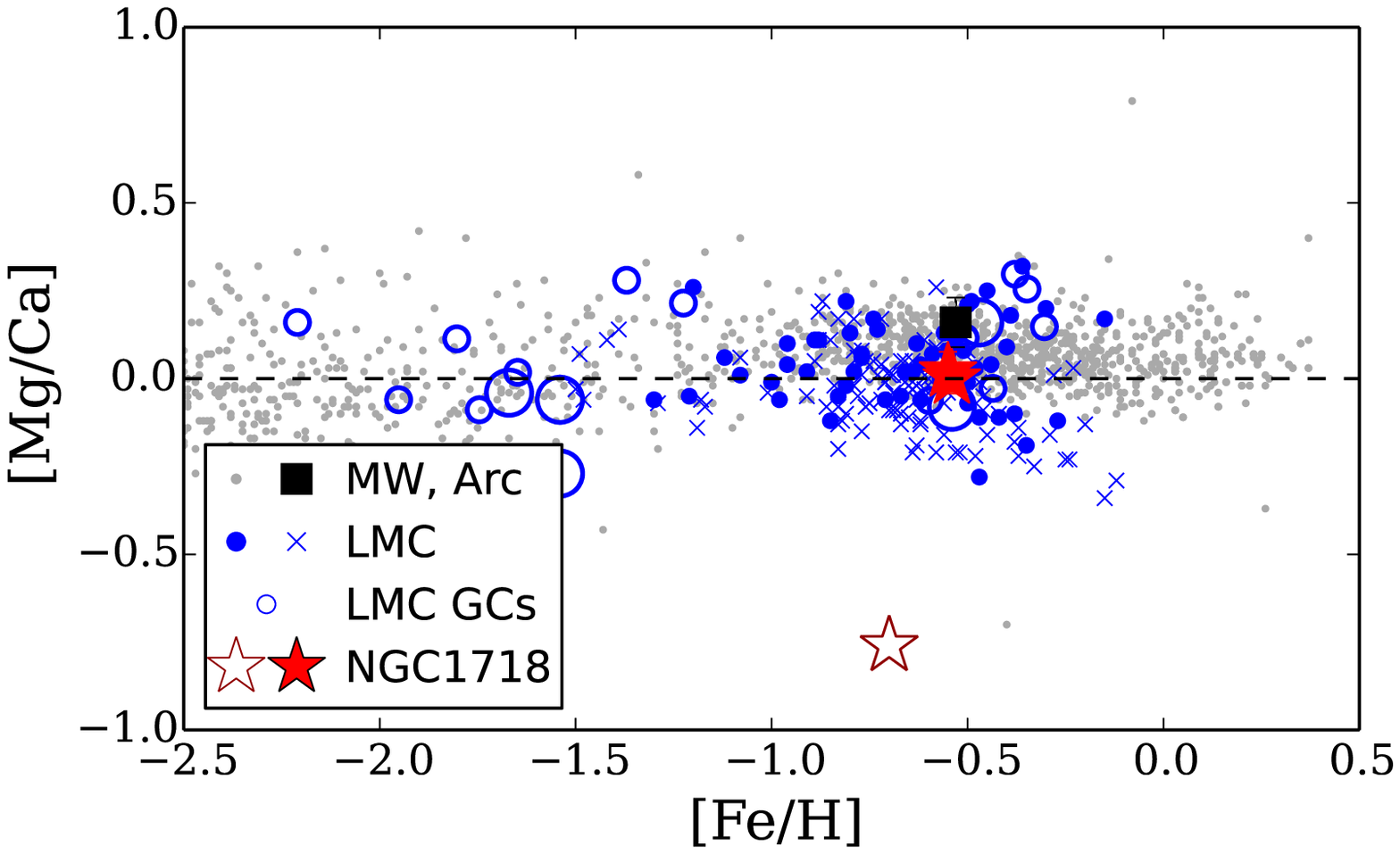}\label{fig:MgCaFe}}
\subfigure{\includegraphics[scale=0.6,trim=0.7in 0in 1.25in 0.0in]{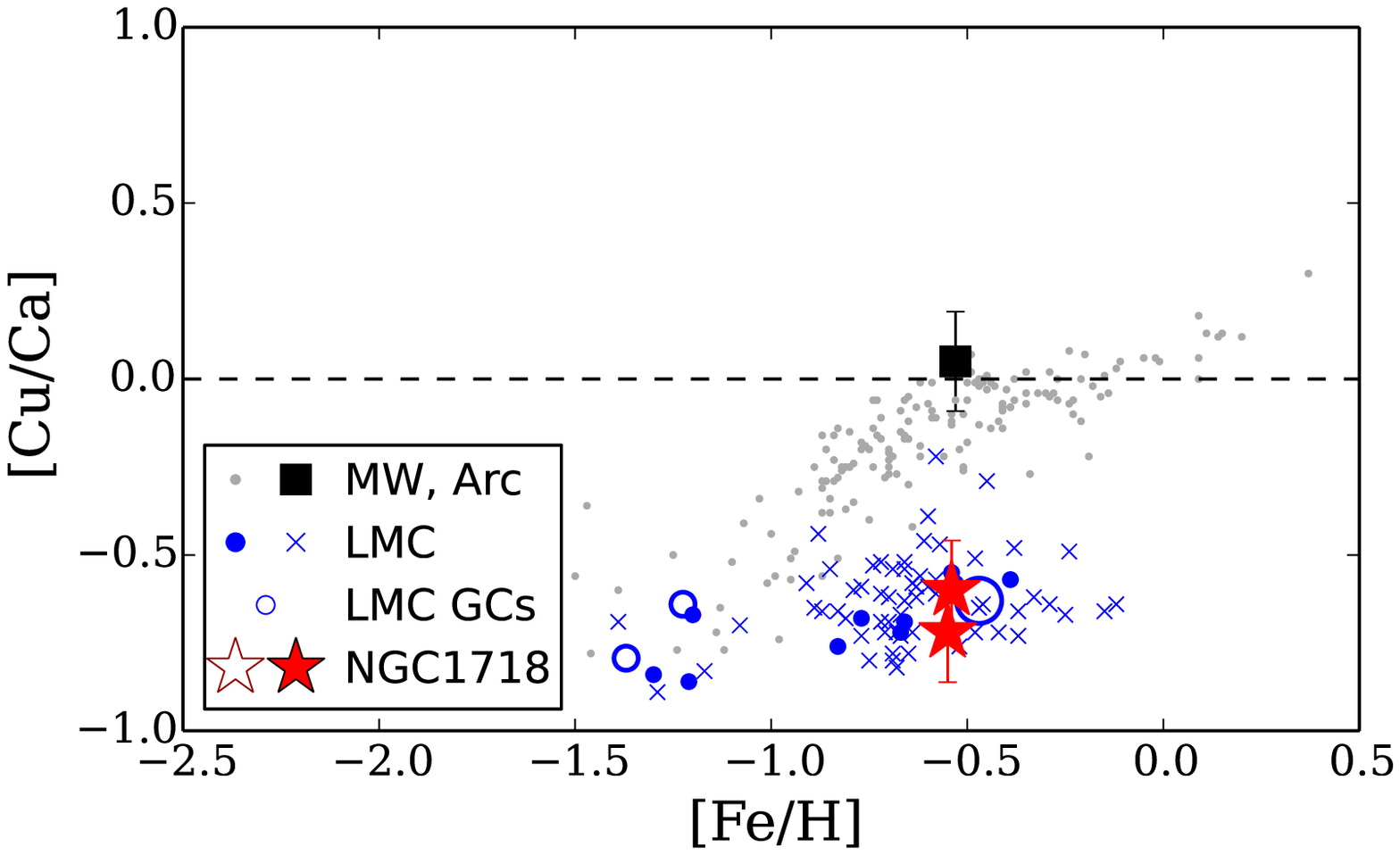}\label{fig:CuCaFe}}
\caption{[Mg/Ca] (left) and [Cu/Ca] (right) ratios in NGC~1718
  compared to MW and LMC field stars and other LMC GCs.  Points are as
  in Figure \ref{fig:NaOfigs}.}\label{fig:MgCafigs}
\end{center}
\end{figure*}

The traditional $\alpha$-elements do not all share the same formation
site.  \citet{McW2013} outline two broad categories of
$\alpha$-elements: {\it hydrostatic} elements like O and Mg, which
form during normal burning phases in massive stars; and {\it
  explosive} elements like Si, Ca, and Ti, which form in Type II
supernovae.  \citet{McW2013} also add Na, Al, and Cu to the list of
hydrostatic elements, since Na and Al are also expected to form during
C and Ne burning \citep{WoosleyWeaver1995}, and Cu is expected to form
from the weak s-process during hydrostatic burning in massive stars.
In the NGC~1718 stars, the Na, Al, and Cu ratios are all lower than
the MW field stars.  Relative contributions of hydrostatic and
explosive elements can be considered by examining ratios, e.g.,
[Mg/Ca] and [Cu/Ca] (see Figure \ref{fig:MgCafigs}).  In the
Sagittarius dwarf spheroidal (Sgr), McWilliam et al. found that the
hydrostatic and explosive elements did not track each other;
specifically, they found a paucity of hydrostatic elements compared to
explosive elements (seen as low [Mg/Ca], for example). They
interpreted these offsets as a signature of a top-light initial mass
function (IMF), i.e., one lacking the most massive stars, which are
expected to produce hydrostatic elements.

\begin{table*}
\centering
\begin{minipage}{165mm}
\begin{center}
\caption{[X/Ca] ratios.\label{table:XCa}}
  \newcolumntype{e}[1]{D{.}{.}{#1}}
  \begin{tabular}{@{}le{4}e{4}e{5}ce{5}e{6}@{}}
  \hline
 & \multicolumn{3}{c}{[X/Ca]} & & \multicolumn{2}{c}{$\Delta$[X/Ca]$^{a}$}  \\
 & \multicolumn{1}{c}{Arcturus} & \multicolumn{1}{c}{NGC~1718-9} & \multicolumn{1}{c}{NGC~1718-26} & & \multicolumn{1}{c}{NGC~1718-9} & \multicolumn{1}{c}{NGC~1718-26} \\
\hline
Cu          &  0.05 & -0.72 & -0.60 & & -0.77 & -0.65 \\
O           &  0.10 & -0.22 & -0.22 & & -0.32 & -0.32 \\
Al          &  0.21 & -0.08 & -0.07 & & -0.29 & -0.28 \\
Na          & -0.07 & -0.22 & -0.29 & & -0.15 & -0.22 \\
Mg          &  0.16 &  0.02 &  0.0  & & -0.14 & -0.16 \\
Si          &  0.10 &  0.02 &  0.02 & & -0.08 & -0.08 \\
\ion{Ti}{1} &  0.07 &  0.0  & -0.05 & & -0.07 & -0.12 \\
\hline
\end{tabular}
\end{center}
\end{minipage}\\
\medskip
\raggedright $^{a}$ Relative to Arcturus.\\
\medskip
\end{table*}

The LMC results are not as straightforward as the Sgr results.  In
their sample of bar stars, \citet{VanderSwaelmen2013} found that the
hydrostatic elements were occasionally lower than the explosive
elements, albeit with a large scatter.  The NGC~1718 stars generally
follow the LMC disk stars: the disk stars have similar [Mg/Fe] and
[Ca/Fe] ratios, such that [Mg/Ca] is roughly solar\footnote{Note that
  with the original \citet{Pompeia2008} results [Mg/Ca] was
  supersolar.  \citet{VanderSwaelmen2013} subsequently revised the
  Pomp\'{e}ia et al. Mg abundances downward---these are the values
  given in the plots.} (see Figure \ref{fig:MgCaFe}).  The bar stars
have slightly lower Mg and higher Ca, which leads to a mildly subsolar
[Mg/Ca].  The offset between hydrostatic and explosive elements is
therefore apparently smaller in the LMC than in Sgr, presumably
because the LMC is more a massive galaxy than Sgr, and can therefore
form more massive stars.  However, [O/Ca] is mildly low in the two
NGC~1718 stars, at $-0.22$.  In NGC~1718 and the LMC field stars the
other hydrostatic elements Na, Al, and Cu are lower than the
explosive elements (see Figure \ref{fig:CuCaFe}). The Sgr results
qualitatively show similar behaviour, i.e. Na, Al, and Cu are lower
than Mg.  This suggests that the LMC is not forming Na, Al, and
Cu as efficiently as Mg and the explosive elements (though note that
the two NGC~1718 stars in this work have higher Na and Al than the
LMC disk stars, which suggests that the offsets in the hydrostatic
elements in not uniform throughout the LMC).

Table \ref{table:XCa} shows [X/Ca] ratios in Arcturus and the NGC~1718
stars for all the explosive and hydrostatic elements considered
here. Compared to Arcturus, the two NGC~1718 stars are deficient in
all hydrostatic elements relative to Ca.  The strongest deficiencies
are in Cu and O, with slightly weaker deficiencies in Na, Mg, and
Al. This could be due to truncation of the high mass end of the IMF
\citep{McW2013} but at a higher mass than for Sgr.  It is unlikely
that low O could be a result of CNO cycle oxygen burning, since the
CNO cycle is expected to reduce oxygen abundances only slightly, by
$\sim 0.05$ dex \citep{LambertRies1981}.  On the other hand, extensive
burning of oxygen should also produce prodigious amounts of Na, which
is not seen.

\subsubsection{Fe-peak Elements}\label{subsubsec:DiscussFePeak}
The Mn abundance ratios in NGC~1718 are subsolar, while [V/Fe] and
[Ni/Fe] are roughly solar.  [V/Fe] is in agreement with other LMC
stars and GCs, through there is a large scatter.  There are very few
[Mn/Fe] measurements in LMC stars; however, NGC~1718 agrees with most
of the GCs that have had Mn determinations.  MW field stars also have
subsolar [Mn/Fe] ratios at $[\rm{Fe/H}]~=~-0.5$.   Lower [Mn/Fe] has
been seen in various dwarf galaxies, notably Sgr; however,
\citep{McW2013} found solar [Mn/Fe] and [V/Fe] in their differential
analyses of three Sgr field stars, and argued that previous reports of
low V and Mn in Sgr stars may result from problems with, e.g., the
adopted stellar temperatures, solar abundances, atomic data, and/or
non-LTE corrections.  All of these issues can be minimized with a
line-by-line differential analysis. The differential analysis in this
work also finds similar [Mn/Fe] and [V/Fe] as MW field stars.

\subsubsection{Neutron Capture Elements: The s-Process}\label{subsubsec:DiscussNeutronCapture}

\begin{figure*}
\begin{center}
\centering
\subfigure{\includegraphics[scale=0.6,trim=1.0in 0in 0.05in 0.0in]{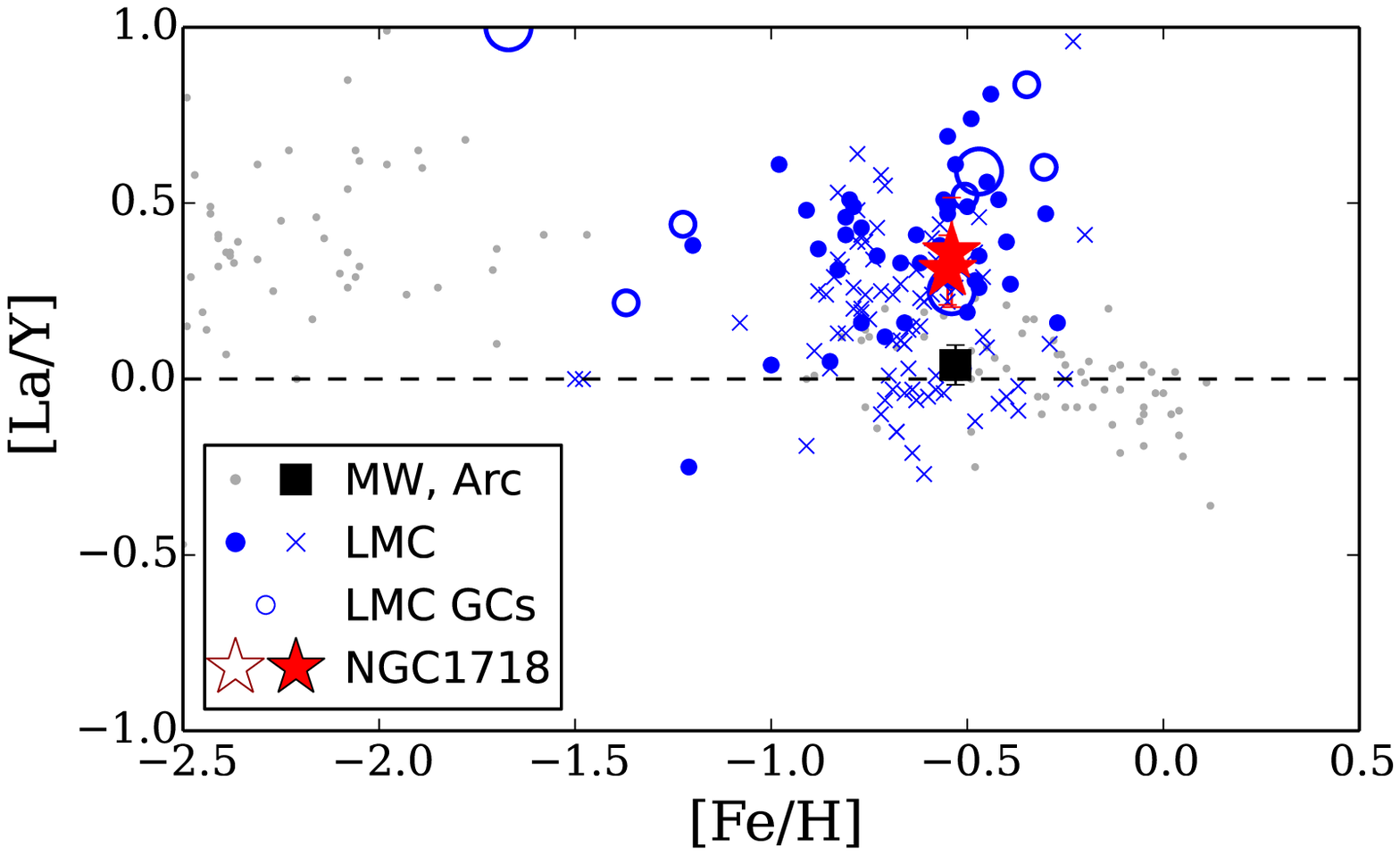}\label{fig:LaY}}
\subfigure{\includegraphics[scale=0.6,trim=0.7in 0in 1.25in 0.0in]{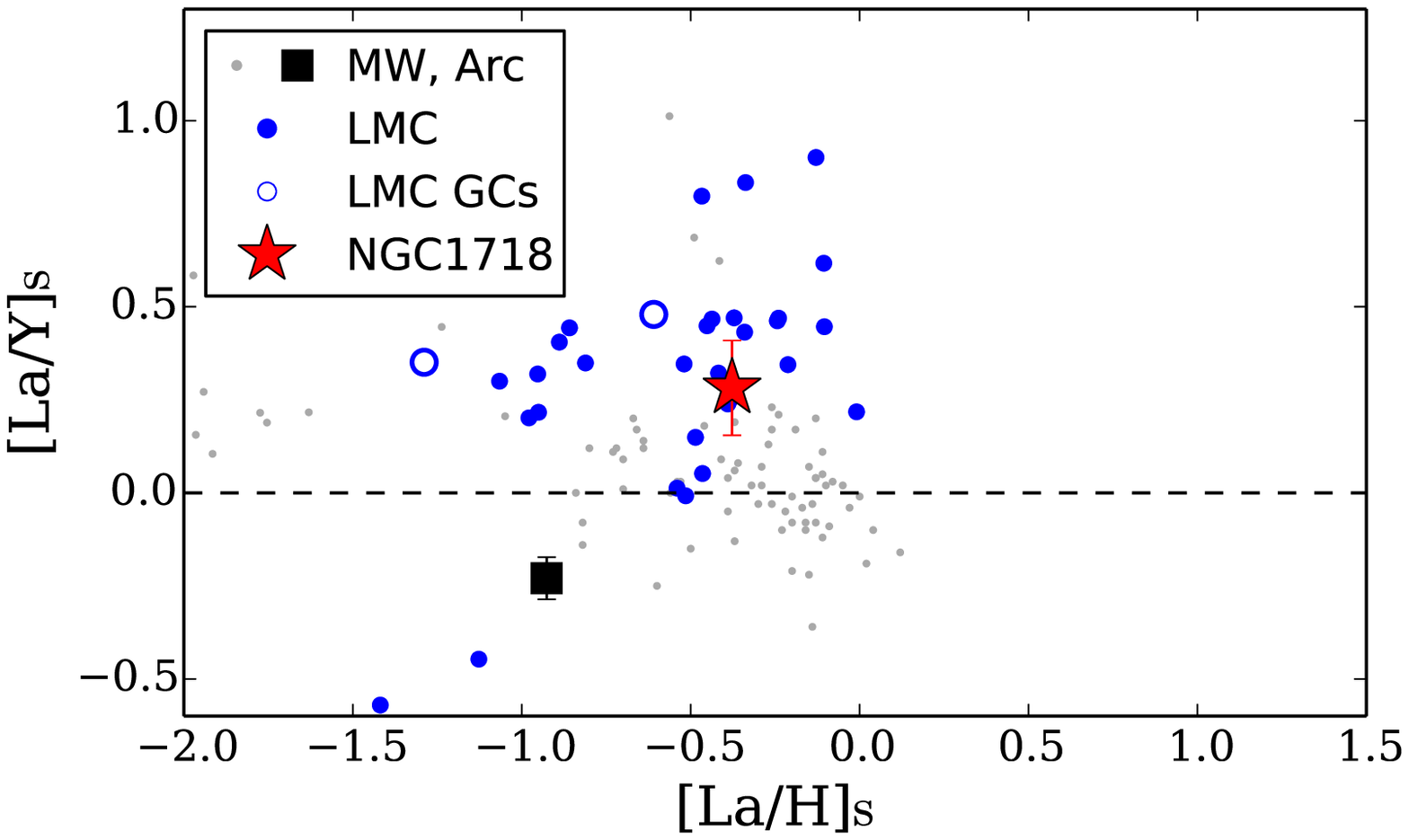}\label{fig:LaHLaY}}
\caption{[La/Y] ratios versus [Fe/H] (left) and [La/H] (right) in NGC~1718
  compared to MW and LMC field stars and other LMC GCs.  Points are as
  in Figure \ref{fig:NaOfigs}, with the addition of MW field stars
  from \citet{Simmerer2004}.  Because their Fe, Y, and La abundances
  are nearly identical, the average of the NGC~1718 stars is shown.
  In the right panel, the abundance ratios have been ``r-process
  subtracted'' to show only contributions from the s-process (see
  \citealt{McW2013}).}\label{fig:LaYFigs}
\end{center}
\end{figure*}

\paragraph{The s-Process}\label{subsec:sDiscussion}
Table \ref{table:Abunds} shows that [La/Fe] is mildly enhanced in
NGC~1718, by $0.2-0.3$ dex compared to Arcturus.  This agrees with the
LMC bar and disk stars and the other GCs, and is consistent with other
dwarf galaxies (e.g.,
\citealt{SHMcW2002,Shetrone2003,Venn2004,McW2013}), which are known to
have excesses of second peak (heavy) s-process elements like La
compared to first peak (light) elements like Y and Zr. Recall that the
main site of the s-process is thought to be AGB stars.  In metal-poor
AGB stars there is a high ratio of neutrons to seed nuclei, which
enables the s-process to create heavier nuclei than in metal-rich AGB
stars.  Stars in dwarf galaxy are thought to receive more s-process
enrichment from metal-poor AGB stars, which leads to a higher [La/Y]
than MW field stars (see Figure \ref{fig:LaY}).  This difference is
not due to any contributions from the r-process: Figure
\ref{fig:LaHLaY} shows s-process only ratios, where the r-process
contributions have been subtracted, following \citet{McW2013}.
Indeed, NGC~1718's [La/Y] ratio is higher than Arcturus and other MW
field stars, in agreement with the LMC stars\footnote{Note that the
  LMC bar stars show a large spread in [La/Y] at a fixed [La/H]---this
  is due to the Y abundances, which are slightly higher than the LMC
  disk stars.  For this reason the bar stars are not included in
  Figure \ref{fig:LaHLaY}.  \citet{VanderSwaelmen2013} confirm that
  this is not likely to be a systematic error, and they suggest that
  the high Y may therefore indicate that the bar stars have retained
  the ejecta of more metal-rich AGB stars.  The scatter in [La/Y]
  suggests that the bar may have experienced inhomogeneous mixing,
  with some environments retaining metal-rich AGB ejecta and others
  experiencing leaky box chemical evolution and retention of
  metal-poor AGB ejecta.  Comparisons with the dilution curves in
  \citet{McW2013} show that this scatter can be reproduced with
  varying yields of La and Y.}
and the Sgr stars from \citet{McW2013}. Zr (another first peak
s-process element) is also lower than La in these stars.

Rb is produced by the s-process (50\% in the sun;
\citealt{Burris2000})---the dominant contributor is thought to be
intermediate-mass AGB stars (e.g., \citealt{Fishlock2014}).
\citet{McW2013} provide a through explanation of Rb's nucleosynthetic
properties.  Their comparison of Rb ($Z=37$) with Zr ($Z=40$) showed
that Rb was underabundant with respect to Zr, as compared to MW field
stars.  This is also seen in NGC~1718, where [Rb/Zr] is $\sim 0.1-0.2$
dex lower than Arcturus. \citet{McW2013} attribute this offset in Sgr
to a lower retention of ejecta from intermediate-mass AGB stars at
high [Fe/H], compared to lower-mass AGB stars.  While this could
indicate a paucity of intermediate-mass AGB stars compared to the MW
(due to, e.g., a top-light IMF), McWilliam et al. also implicate
``leaky box'' chemical evolution as a means of selectively enriching a
galaxy with low-mass AGB stars.

The LMC stars therefore generally show the excesses in heavy
s-process elements that are typically seen in dwarf galaxies.  The
LMC's offsets from the MW are not as significant as those seen in Sgr,
which is consistent with the LMC being a more massive galaxy.

\paragraph{The r-Process}\label{subsec:r}
Eu is primarily an r-process element in the Sun \citep{Burris2000}.
As can be seen in Table \ref{table:Abunds}, NGC~1718-9 and -26 show no
enhancements over Arcturus or the other MW field stars.  This suggests
that the LMC is not enhanced in r-process elements, relative to MW
stars at the same [Fe/H].  The LMC disk and bar stars do seem to be
slightly enhanced in [Eu/Fe]; \citet{VanderSwaelmen2013} argue that
this is not a systematic offset since their Arcturus abundance is in
agreement with the MW field stars.  However, their Arcturus Eu
abundance is $\sim 0.1$ dex higher than the value derived in this
analysis.  If the \citet{VanderSwaelmen2013} [Eu/Fe] ratios are all
lowered by 0.1 dex, most of the stars are in agreement with the MW
field stars.

\begin{figure*}
\begin{center}
\centering
\subfigure{\includegraphics[scale=0.6,trim=1.0in 0in 0.05in 0.0in]{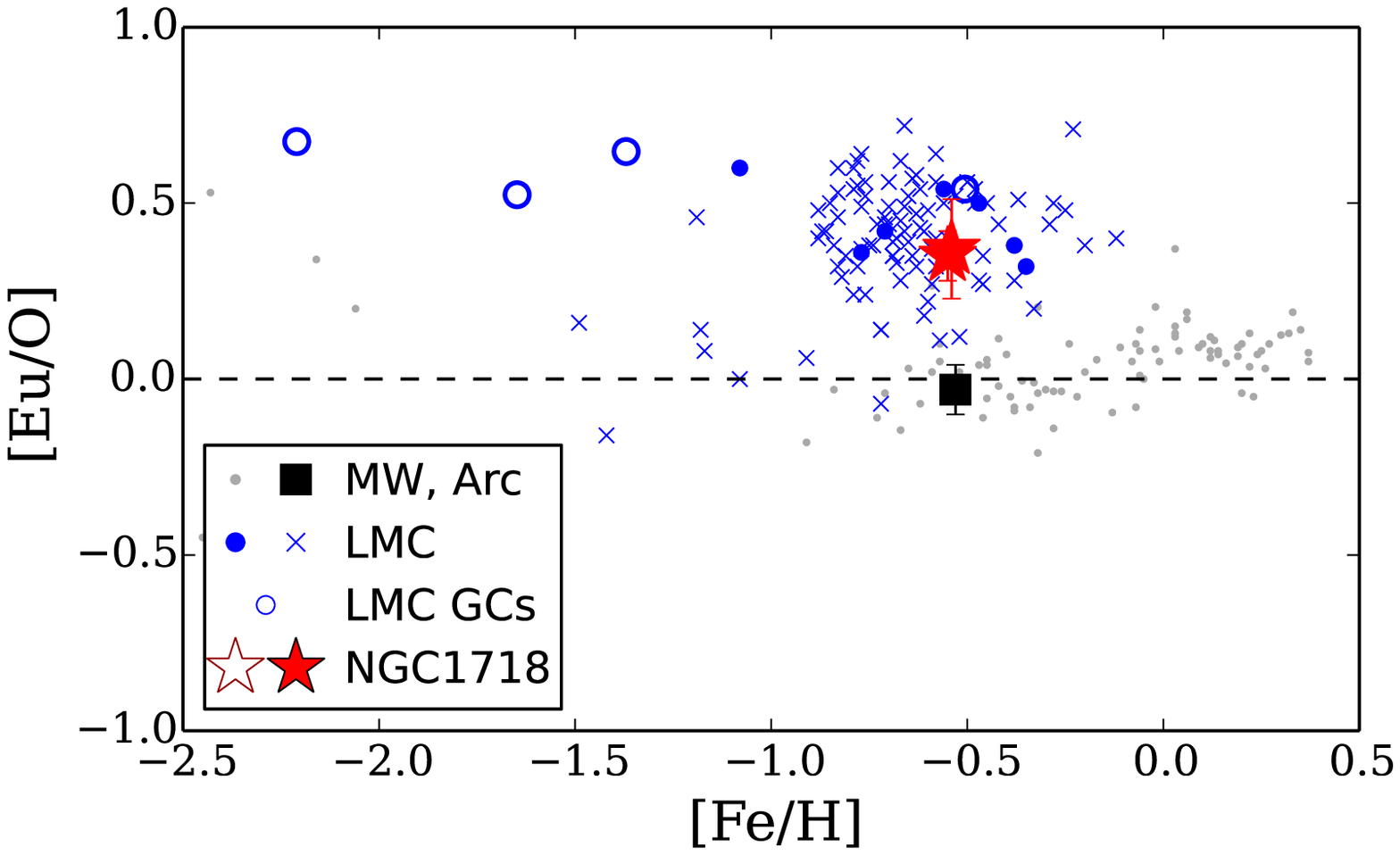}\label{fig:EuO}}
\subfigure{\includegraphics[scale=0.6,trim=0.7in 0in 1.25in 0.0in]{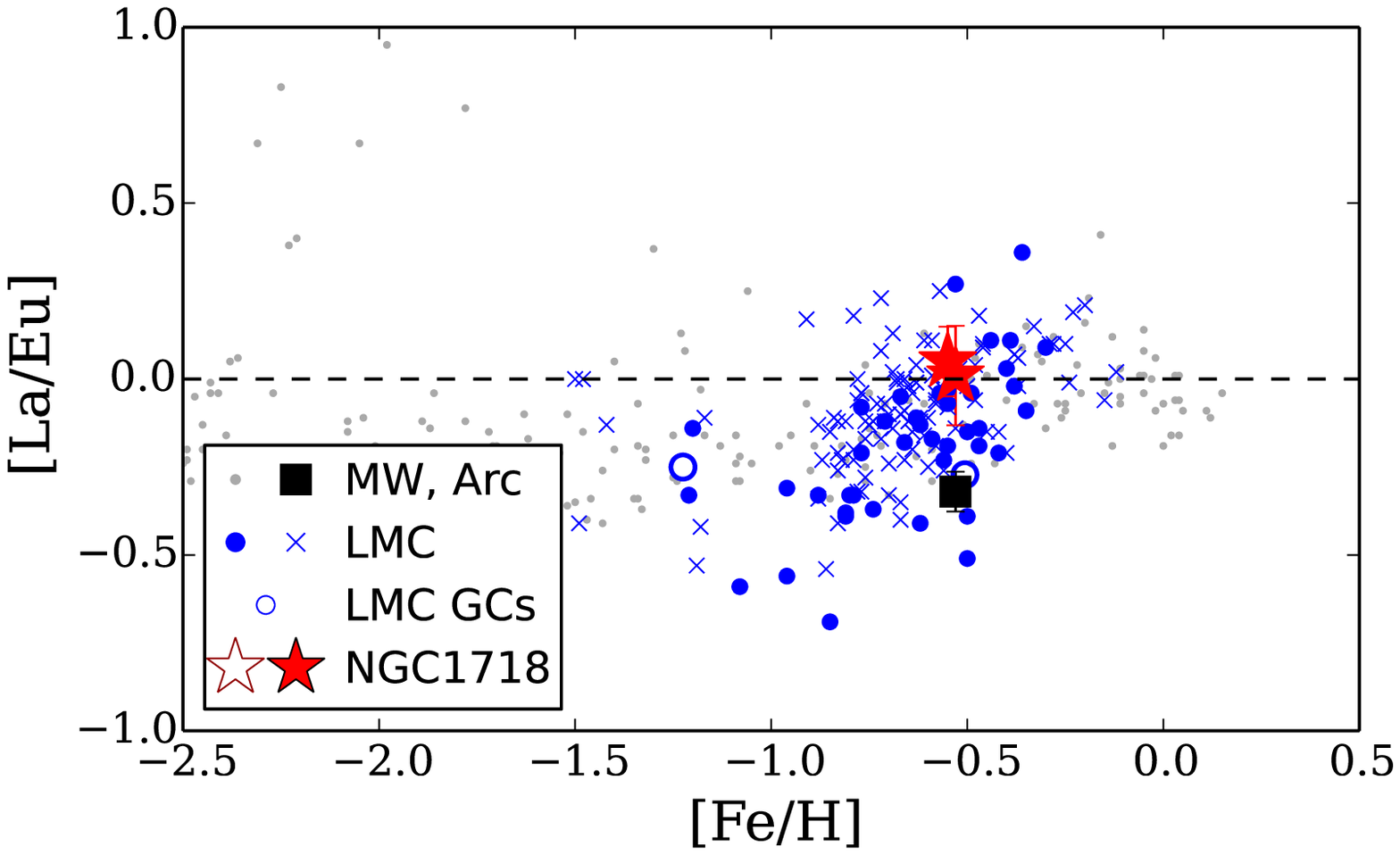}\label{fig:LaEu}}
\caption{[Eu/O] (left) and [La/Eu] (right) ratios in NGC~1718
  compared to MW and LMC field stars and other LMC GCs.  Points are as
  in Figure \ref{fig:NaOfigs}.}\label{fig:NeutronCaptureFigs}
\end{center}
\end{figure*}

Although the site of the r-process is not well constrained, core
collapse supernovae \citep{QianWasserburg2008} and neutron star
mergers \citep{TsujimotoShigeyama2014} are leading candidates.  Both
of these candidate sites fall at the end stages of massive star
evolution.  Recall that the hydrostatic elements O, Mg, etc. are also
thought to form in massive stars (the progenitors of core collapse
supernovae).  Figure \ref{fig:EuO} compares Eu to the hydrostatic
element O, and demonstrates that the LMC stars are enhanced in Eu
relative to O.  As discussed in Section \ref{subsubsec:DiscussAlpha},
this is driven by deficiencies in the O abundances.  Similar offsets
are seen in the Sgr stars analyzed by \citet{McW2013}.  In their
framework, this indicates that the r-process occurs in stars with
lower initial masses than the stars that produce the hydrostatic
elements, indicating that Sgr and the LMC are missing the highest mass
stars.  However, regardless of the cause of the O deficiencies, the
high [Eu/O] ratios relative to the MW indicate that the
nucleosynthesis of r-process elements cannot occur in the same stars
that produce most of the hydrostatic elements; otherwise, the ratios
would be the same between the LMC and the MW.

\paragraph{The s- vs. the r-Process}\label{subsec:svsr}
Figure \ref{fig:LaEu} shows [La/Eu] ratios. The
ratio of [La/Eu] gives a rough indication of the relative
contributions from the s-process vs. the r-process.  As AGB stars
evolve, [La/Eu] gradually rises over time.  Relative to Arcturus, the
NGC~1718 stars have high [La/Eu], similar to the Sgr stars in
\citet{McW2013}.  Since there is no convincing offset between the LMC
and Arcturus in r-process contributions, the offset in [La/Eu] is
driven entirely by the s-process; specifically, the LMC is enhanced
in heavy s-process elements.

\section{Conclusion}\label{sec:Conclusion}
This paper has presented detailed chemical abundances of two stars in
the intermediate-age ($\sim 2$ Gyr) LMC cluster NGC~1718.  The
analysis was performed {\it differentially} with respect to Arcturus.
This differential analysis drastically reduces the systematic
uncertainties that will occur in a typical abundance analysis; as a
result, these abundances have high precision.  Although
two stars represent a small fraction of the stars in the cluster,
several important findings are evident from this analysis.

\begin{enumerate}
\item The two stars have nearly identical heliocentric radial
  velocities and abundance ratios, indicating that these stars are
  both cluster members.

\item The abundances of these stars are similar to LMC disk and bar
  field stars, suggesting a) that the abundances of these stars
  represent the primordial abundances of the giant molecular cloud
  from which NGC~1718 formed, and b) that these stars are useful for
  investigating the chemical evolution of the LMC.

\item The cluster is found to have an average metallicity of
  $[\rm{Fe/H}] = -0.55\pm 0.01$, a value which is consistent with the
  age-metallicity relationship for the LMC.

\item The [Mg/Fe] ratios of the two stars are similar to LMC field
  stars, demonstrating that the low [Mg/Fe] derived by \citet{Colucci2012}
  from integrated light spectroscopy cannot be due to a low primordial
  Mg abundance in NGC~1718's birth environment.  Although no signs of
  Na, O, Mg, and/or Al spreads are seen between the two stars, this
  analysis cannot rule out the possibility that the low integrated
  [Mg/Fe] is due to star-to-star variations within the cluster,
  although this seems unlikely given the low mass and high metallicity
  of this cluster.

\item The [X/Fe] ratios of the explosive elements Si, Ca, and Ti are
  moderately enhanced, at $\sim 0.1$ dex, which is slightly low
  compared to MW field stars.  Relative to Ca, NGC~1718 shows
  deficiencies in elements which form via hydrostatic burning in
  massive stars; the deficiencies in Na, Mg, and Al are moderate, while
  those in O and Cu are strong.

\item The iron-peak elements V and Mn are roughly consistent with MW
  and LMC field stars.

\item The s-process abundances, specifically the [La/Y] ratios,
  indicate that the LMC disk was enriched by metal-poor AGB stars.

\item NGC~1718 shows no enhancement in the r-process relative to MW
  field stars of similar metallicity.
\end{enumerate}

The two NGC~1718 stars in this analysis track the chemical properties
of the LMC field stars.  With only two stars, this analysis cannot
rule out the possibility of multiple populations in other cluster
stars, and NGC~1718 therefore remains an interesting cluster for
future study.  Analyses of more stars at various evolutionary stages
would be necessary to rule out any signs of abundance variations
within the cluster. When NGC~1718 is utilized to probe the chemical
evolution of the LMC, its high precision, differential abundances
indicate that LMC's chemical abundances can be explained by a ``top
light'' IMF that is missing the highest mass stars, as was proposed
for Sgr by \citet{McW2013}.

\section*{Acknowledgments}
The authors thank the anonymous referee for comments that
improved the manuscript.  CMS and GW acknowledge funding from the
Kenilworth Foundation. This publication makes use of data products
from the Two Micron All Sky Survey, which is a joint project of the
University of Massachusetts and the Infrared Processing and Analysis
Center/California Institute of Technology, funded by the National
Aeronautics and Space Administration and the National Science
Foundation.

\footnotesize{

}

\begin{onecolumn}
\normalsize
\appendix
\section{Error Analysis}\label{appendix:Errors}

In this work the uncertainties in the abundance measurements are
computed using equations similar to those employed by
\citet{McWilliam1995} and \citet{McW2013}.  The uncertainty in the
adopted T$_{\rm eff}$ is based on the scatter in the slope of the
plot of Fe~I abundance with EP, plus the uncertainty of 29K in the
temperature of Arcturus (used in the differential abundance analysis),
which was added in quadrature. The resultant 1$\sigma$ T$_{\rm eff}$
uncertainty of 51K (see Table~\ref{table:varcovar}) is obtained for
both stars.  The adopted model atmosphere surface gravities were taken
from BaSTI isochrones (Pietrinferni et al. 2004), with an age of 2~Gyr
and appropriate metallicity, based on the adopted temperature (see
Section \ref{sec:AtmParams}); the temperature uncertainty and an age
uncertainty of 1~Gyr were then used to estimate 1$\sigma$ on log\,g of
0.089 dex cm s$^{-2}$.  This method for adopting gravity resulted in a
significant covariance between T$_{\rm eff}$ and log\,g, for which a
temperature-gravity covariance, $\sigma_{Tg}$, of 4.42 is found. The
computed variances and covariance for the remaining atmospheric
parameters are found following \citet{McW2013}, and are shown in
Table~\ref{table:varcovar}.

\begin{table*}
\centering
\begin{minipage}{165mm}
\begin{center}
\caption{Atmosphere Parameter Variances and Covariances.\label{table:varcovar}}
  \newcolumntype{d}[1]{D{,}{\pm}{#1}}
  \newcolumntype{e}[1]{D{.}{.}{#1}}
  \begin{tabular}{@{}ce{10}@{}}
\hline
Parameters   & \multicolumn{1}{c}{Variance/covariance} \\
  \hline
$\sigma_{T}$        &  51.0       \\
$\sigma_{g}$        &  0.086      \\
$\sigma_{\xi}$      &  0.06       \\
$\sigma_{[\rm{M/H}]}$     &  0.04       \\
                   &               \\
$\sigma_{Tg}$       &  4.42       \\
$\sigma_{T[\rm{M/H}]}$    & -1.56   \\
$\sigma_{[\rm{M/H}]\xi}$  & -0.02    \\
$\sigma_{g\xi}$  &  0.0^{a}     \\
$\sigma_{T\xi}$  &  0.0^{a}     \\
\hline
\end{tabular}
\end{center}
\end{minipage}\\
\medskip
$^{a}$ The computed covariance values are sufficiently small that zero
is adopted.\\
\end{table*}

The uncertainty in an average abundance ratio,
$\overline{\varepsilon_1/\varepsilon_2}$, was found using an expansion
similar to Equation A16 in \citet{McWilliam1995}:

\begin{eqnarray}\label{eqn:A1}
 \sigma (\overline{\varepsilon_{1}/\varepsilon_{2}})^2  = \sigma_r(\overline{\varepsilon_1/\varepsilon_2} )^2 +
                      \left(\frac{\partial \overline{\varepsilon_1/\varepsilon_2}}{\partial T}\right)^2   \sigma_{T}^2  +
                      \left(\frac{\partial \overline{\varepsilon_1/\varepsilon_2}}{\partial g}\right)^2   \sigma_{g}^2 +
                      \left(\frac{\partial \overline{\varepsilon_1/\varepsilon_2}}{\partial \xi}\right)^2    \sigma_{\xi}^2 +  
                      \left(\frac{\partial \overline{\varepsilon_1/\varepsilon_2}}{\partial [M/H]}\right)^2  \sigma_{[M/H]}^2 +  \nonumber \\
             2 \left[ \left(\frac{\partial \overline{\varepsilon_1/\varepsilon_2}}{\partial T}\right)  
                      \left(\frac{\partial \overline{\varepsilon_1/\varepsilon_2}}{\partial g}\right) \sigma_{Tg}  +  
                           \left(\frac{\partial \overline{\varepsilon_1/\varepsilon_2}}{\partial T}\right)                     
                           \left(\frac{\partial \overline{\varepsilon_1/\varepsilon_2}}{\partial \xi}\right)  \sigma_{T\xi} +   
                           \left(\frac{\partial \overline{\varepsilon_1/\varepsilon_2}}{\partial g}\right)                     
                           \left(\frac{\partial \overline{\varepsilon_1/\varepsilon_2}}{\partial \xi}\right)  \sigma_{g\xi} +  
                           \left(\frac{\partial \overline{\varepsilon_1/\varepsilon_2}}{\partial [M/H]}\right)                
                           \left(\frac{\partial \overline{\varepsilon_1/\varepsilon_2}}{\partial T}\right) \sigma_{T[M/H]} \right]
\end{eqnarray}

The abundance sensitivity to atmospheric parameters for the lines in
Section \ref{sec:LineList} was computed for NGC~1718-26 and appears in
Table~\ref{table:dabunddpar}; these gradients were subsequently
employed with the variances, covariances, and Equation \ref{eqn:A1} to
determine abundance ratio uncertainties. The final uncertainties in
[X/H], [X/\ion{Fe}{1}] and [X/\ion{Fe}{2}] appear in
Table~\ref{table:finalerrors}.

\begin{table*}
\centering
\begin{minipage}{165mm}
\begin{center}
\caption{Abundance Sensitivity to Atmosphere Parameters.\label{table:dabunddpar}}
  \newcolumntype{d}[1]{D{,}{\pm}{#1}}
  \newcolumntype{e}[1]{D{.}{.}{#1}}
  \begin{tabular}{@{}lccccccccccc@{}}
  \hline
 & \multicolumn{2}{c}{$\Delta$ T$_{\rm eff}$ (K)} & & \multicolumn{2}{c}{$\Delta$ log\,g (dex)} & & \multicolumn{2}{c}{$\Delta$ $\xi$ (km s$^{-1}$)} & & \multicolumn{2}{c}{$\Delta$ [M/H] (dex)} \\
 & $+$50 & $-$50 & & $+$0.2 & $-$0.2 & & $+$0.3 & $-$0.3 & & $+$0.1 & $-$0.1 \\
\hline
Fe~I   &  $-$0.03 &  $+$0.03   &  &        $+$0.04 &  $-$0.07  & &           $-$0.09 &  $+$0.11   & &          $+$0.03 & $-$0.04  \\
Fe~II  &  $-$0.12 &  $+$0.12   &  &        $+$0.07 &  $-$0.16  & &           $-$0.05 &  $+$0.07   & &          $+$0.05 & $-$0.08  \\
{[O~I]} &  $+$0.01 & $-$0.02   &  &        $+$0.08 &  $-$0.09  & &           $-$0.03 &  $+$0.04   & &          $+$0.04 & $-$0.05  \\
Na~I   &  $+$0.04 &  $-$0.04   &  &        $-$0.02 &  $-$0.00  & &           $-$0.08 &  $+$0.09   & &          $+$0.01 & $-$0.01  \\
Mg~I   &  $-$0.04 &  $+$0.04   &  &        $+$0.01 &  $-$0.05  & &           $-$0.03 &  $+$0.03   & &          $+$0.02 & $-$0.03  \\
Al~I   &  $+$0.02 &  $-$0.02   &  &        $+$0.00 &  $-$0.01  & &           $-$0.05 &  $+$0.06   & &          $+$0.01 & $-$0.02  \\
Si~I   &  $-$0.07 &  $+$0.07   &  &        $+$0.03 &  $-$0.09  & &           $-$0.04 &  $+$0.05   & &          $+$0.03 & $-$0.05  \\
Ca~I   &  $+$0.05 &  $-$0.05   &  &        $-$0.00 &  $-$0.01  & &           $-$0.07 &  $+$0.10   & &          $+$0.01 & $-$0.01  \\
Ti~I   &  $+$0.05 &  $-$0.05   &  &        $+$0.03 &  $-$0.03  & &           $-$0.11 &  $+$0.14   & &          $+$0.03 & $-$0.03  \\
Ti~II  &  $-$0.04 &  $+$0.04   &  &        $+$0.07 &  $-$0.11  & &           $-$0.04 &  $+$0.05   & &          $+$0.04 & $-$0.05  \\
V~I    &  $+$0.08 &  $-$0.03   &  &        $+$0.07 &  $-$0.03  & &           $-$0.16 &  $+$0.22   & &          $+$0.06 & $-$0.02  \\
Mn~I   &  $-$0.01 &  $+$0.00   &  &        $+$0.04 &  $-$0.07  & &           $-$0.13 &  $+$0.14   & &          $+$0.03 & $-$0.04  \\
Ni~I   &  $-$0.03 &  $+$0.03   &  &        $+$0.05 &  $-$0.08  & &           $-$0.09 &  $+$0.11   & &          $+$0.03 & $-$0.05  \\
Cu~I   &  $-$0.01 &  $+$0.00   &  &        $+$0.06 &  $-$0.09  & &           $-$0.11 &  $+$0.13   & &          $+$0.04 & $-$0.05  \\
Rb~I   &  $+$0.06 &  $-$0.06   &  &        $+$0.01 &  $-$0.00  & &           $-$0.01 &  $+$0.01   & &          $+$0.01 & $-$0.01  \\
Y~II   &  $-$0.02 &  $+$0.02   &  &        $+$0.07 &  $-$0.09  & &           $-$0.02 &  $+$0.02   & &          $+$0.04 & $-$0.05  \\
Zr~I   &  $+$0.06 &  $-$0.07   &  &        $+$0.03 &  $-$0.03  & &           $-$0.21 &  $+$0.26   & &          $+$0.03 & $-$0.04  \\
La~II  &  $+$0.01 &  $-$0.02   &  &        $+$0.08 &  $-$0.09  & &           $-$0.03 &  $+$0.04   & &          $+$0.04 & $-$0.05  \\
Eu~II  &  $-$0.01 &  $+$0.00   &  &        $+$0.07 &  $-$0.09  & &           $-$0.03 &  $+$0.03   & &          $+$0.04 & $-$0.05  \\
\hline
\end{tabular}
\end{center}
\end{minipage}\\
\medskip
\end{table*}

It is notable that the systematic uncertainty on [\ion{Fe}{1}/H] in
Table~\ref{table:finalerrors} is quite small; this is apparently due
to the covariance between the effects of T$_{\rm eff}$ and $\log g$ on
[\ion{Fe}{1}/H]: for the \ion{Fe}{1} and II lines in these cool stars
an increase in temperature results in a slightly lower \ion{Fe}{1}
abundance (due to excitation); however, this change in the \ion{Fe}{1}
abundance is compensated-for by the corresponding change in $\log
g$. The same temperature change results in a relatively large decrease
in the derived \ion{Fe}{2} abundance, presumably due to an increase in
Fe ionization. Table~\ref{table:finalerrors} also shows that for most
abundance ratios [X/\ion{Fe}{1}] exhibits smaller systematic
uncertainty than [X/\ion{Fe}{2}], even for ionized metals and
[\ion{O}{1}] lines; thus, the [X/\ion{Fe}{1}] abundance ratios provide
the most reliable estimate.

Note that the implied systematic errors in abundance ratios are only
as good as the input variances, co-variances and abundance sensitivity
to atmosphere parameters. For an element ratio where the actual
systematic abundance sensitivity is small, small errors in the
adopted inputs to Equation~\ref{eqn:A1} can result in a negative
computed variance,  implying unrealistic, or imaginary, 1$\sigma$
uncertainties, when they are simply close to zero; in this case, the
uncertainty in the ratio is dominated by the random error component.

\begin{table*}
\centering
\begin{minipage}{165mm}
\begin{center}
\caption{Abundance Ratio Uncertainties.\label{table:finalerrors}}
  \newcolumntype{d}[1]{D{,}{\pm}{#1}}
  \newcolumntype{e}[1]{D{.}{.}{#1}}
  \begin{tabular}{@{}lccce{6}ce{8}@{}}
  \hline
& \multicolumn{3}{c}{Atmosphere Uncertainties} & & & \\
Ion & $\sigma$[X/H] & $\sigma$[X/Fe\,I] & $\sigma$[X/Fe\,II] & \multicolumn{1}{c}{$\sigma_{\rm{rand}}[X/H]^{a}$} & $\sigma_{\rm total}$[X/Fe\,I] & \multicolumn{1}{c}{$\sigma_{\rm total}$[X/Fe\,II]} \\
\hline
Fe~I   &     0.009   &         ...    &     0.066   &  0.01       & 0.01$^{b}$ &  \multicolumn{1}{c}{...}  \\
Fe~II  &     0.073   &        0.066   &      ...    &  0.01       & ...      &    0.07^{b}  \\
{[O~I]} &    0.053   &        0.059   &     0.125   &  0.07^{c}    & 0.09    &    0.14      \\
Na~I   &     0.036   &        0.044   &     0.110   &  0.07       & 0.08     &    0.13      \\
Mg~I   &     0.028   &        0.021   &     0.046   &  0.04       & 0.05     &    0.06      \\
Al~I   &     0.023   &        0.030   &     0.096   &  0.07       & 0.08     &    0.12      \\
Si~I   &     0.046   &        0.039   &     0.028   &  0.03       & 0.05     &    0.04      \\
Ca~I   &     0.053   &        0.060   &     0.126   &  0.10       & 0.12     &    0.16      \\
Ti~I   &     0.064   &        0.071   &     0.137   &  0.03       & 0.08     &    0.14      \\
Ti~II  &     0.009   &        0.006   &     0.071   &  0.10       & 0.10     &    0.12      \\
V~I    &     0.078   &        0.085   &     0.150   &  0.08       & 0.12     &    0.17      \\
Mn~I   &     0.020   &        0.026   &     0.092   &  0.12       & 0.12     &    0.15      \\
Ni~I   &     0.007   &        0.004   &     0.070   &  0.05       & 0.05     &    0.09      \\
Cu~I   &     0.028   &        0.034   &     0.100   &  0.10^{c}    & 0.11     &    0.14      \\
Rb~I   &     0.063   &        0.070   &     0.136   &  0.09       & 0.11     &    0.16      \\
Y~II   &     0.016   &        0.021   &     0.087   &  0.08       & 0.08     &    0.12      \\
Zr~I   &     0.079   &        0.086   &     0.152   &  0.06       & 0.10     &    0.16      \\
La~II  &     0.053   &        0.059   &     0.125   &  0.07       & 0.09     &    0.14      \\
Eu~II  &     0.031   &        0.036   &     0.102   &  0.05^{c}    & 0.06     &    0.11      \\
\hline
\end{tabular}
\end{center}
\end{minipage}\\
\medskip
\raggedright {\bf Notes: }  $^{a}$ For species with more than one measured line, random abundance errors, due to EW
                  uncertainties, were adopted from the 1$\sigma$ dispersion about the mean abundances.\\
               $^{b}$ Instead of $\sigma_{\rm total}$[X/Fe], $\sigma_{\rm total}$[Fe/H] ratios are provided.\\
               $^{c}$ Random abundance errors for O, Cu and Eu were computed including estimates of the 1$\sigma$
                  SS measurement uncertainties.\\
\end{table*}

\end{onecolumn}


\begin{thebibliography}{99}
\bibitem[Alonso et al.(1999)] {Alonso1999} Alonso, A., Arribas, S., \&
  Mart\'{i}nez-Roger, C. 1999, \aaps, 140, 261
\bibitem[Asplund et al.(2009)] {Asplund2009} Asplund, M., Grevesse,
N., Sauval, J.A., \& Scott, P. 2009, \araa, 47, 481
\bibitem[Baumgardt et al.(2013)] {Baumgardt2013} Baumgardt, H.,
  Parmentier, G., Anders, P., \& Grebel, E.K. 2013, \mnras, 430, 676
\bibitem[Bensby et al.(2005)] {Bensby2005} Bensby, T., Feltzing, S.,
  Lundstr\"{o}m, I., \& Ilyin, I. 2005, \aap, 433, 185 
\bibitem[Brocato et al.(2001)] {Brocato2001} Brocato, E., Di Carlo,
  E., \& Menna, G. 2001, \aap, 374, 523
\bibitem[Burris et al.(2000)] {Burris2000} Burris, D.L., Pilachowski,
C.A., Armandroff, T.E., et al. 2000, \apj, 544, 302
\bibitem[Buzzoni et al.(2010)] {Buzzoni2010} Buzzoni, A., Patelli, L.,
  Bellazzini, M., Fusi Pecci, F., \& Oliva, E. 2010, \mnras, 403, 1592
\bibitem[Carretta et al.(2009)] {Carretta2009} Carretta, E.,
Bragaglia, A., Gratton, R., \& Lucatello, S. 2009, \aap, 505, 139
\bibitem[Castelli \& Kurucz(2004)] {KuruczModelAtmRef} Castelli, F. \&
Kurucz, R. L. 2004, in IAU Symp. 210, Modelling of Stellar
Atmospheres, ed. N. Piskunov, W.W. Weiss, \& D. F. Gray (San
Francisco: ASP), A20
\bibitem[Colucci et al.(2011)] {Colucci2011} Colucci, J.E.,
Bernstein, R.A., Cameron, S.A., \& McWilliam, A. 2011, \apj, 735, 55
\bibitem[Colucci et al.(2012)] {Colucci2012} Colucci, J.E., Bernstein,
R.A., Cameron, S.A., \& McWilliam, A. 2012, \apj, 746, 29
\bibitem[Colucci et al.(2013)] {Colucci2013} Colucci, J.E., Duran,
M.F., Bernstein, R.A., \& McWilliam, A. 2013, \apjl, 773, 36
\bibitem[Colucci et al.(2014)] {Colucci2014} Colucci, J.E., Bernstein,
  R.A., \& Cohen, J.G. 2014, \apj, 797, 116
\bibitem[Colucci et al.(2016)] {Colucci2016} Colucci, J.E., Bernstein,
  R.A., \& McWilliam, A. 2016, \apj, {\it in press}, arXiv:1611.02734
\bibitem[D'Orazi et al.(2014)] {D'Orazi2014} D'Orazi, V., Angelou,
  G.C., Gratton, R.G., et al. 2014, \apj, 791, 39
\bibitem[D'Orazi et al.(2015)] {D'Orazi2015} D'Orazi, V., Gratton,
  R.G., Angelou, G.C., et al. 2015, \mnras, 449, 4038
\bibitem[Elson \& Fall(1988)] {ElsonFall1988} Elson, R.A. \& Fall,
  S.M. 1988, \aj, 96, 1383
\bibitem[Fishlock et al.(2014)] {Fishlock2014} Fishlock, C.K.,
  Karakas, A.I., Lugaro, M., \& Yong, D. 2014, \apj, 797, 44
\bibitem[Fulbright et al.(2006)] {Fulbright2006} Fulbright, J.P.,
  McWilliam, A., \& Rich, R.M. 2006, \apj, 636, 821
\bibitem[Fulbright et al.(2007)] {Fulbright2007} Fulbright, J.P.,
  McWilliam, A., \& Rich, R.M. 2007, \apj, 661, 1152
\bibitem[Gonz\'{a}lez Hern\'{a}ndez \& Bonifacio(2009)]
  {GonzalezHernandezBonifacio2009} Gonz\'{a}lez Hern\'{a}ndez, J.I. \&
  Bonifacio, P. 2009, \aap, 497, 497
\bibitem[Gratton et al.(2012)] {Gratton2012} Gratton, R.G., Carretta,
  E., \& Bragaglia, A. 2012, {\it The Astrononmy and Astrophysics
    Review}, 20, 50
\bibitem[Grocholski et al.(2006)] {Grocholski2006} Grocholski, A.J.,
  Cole, A.A., Sarajedini, A., Geisler, D., \& Smith, V.V. 2006, \aj,
  132, 1630
\bibitem[Hendricks et al.(2016)] {Hendricks2016} Hendricks, B.,
  Boeche, C., Johnson, C.I., et al. 2016, \aap, 585, A86
\bibitem[Hinkle et al.(2003)] {Hinkle2003} Hinkle, K., Wallace, L.,
Livingston, W., Ayres, T., Harmer, D., \& Valenti, J. 2003, 	
in \textit{The Future of Cool-Star Astrophysics: 12th Cambridge
Workshop on Cool Stars, Stellar Systems, and the Sun (2001 July 30 -
August 3)}, eds. A. Brown, G.M. Harper, and T.R. Ayres, (University of
Colorado), 851
\bibitem[Hollyhead et al.(2016)] {Hollyhead2016} Hollyhead, K.,
  Kacharov, N., Lardo, C., et al. 2016, \mnras, arXiv:1609.01302
\bibitem[Johnson et al.(2006)] {Johnson2006} Johnson, J.A., Ivans,
I.I., \& Stetson, P.B. 2006, \apj, 640, 801
\bibitem[Kelson(2003)] {Kelson2003} Kelson, D.D. 2003, \pasp, 115, 688
\bibitem[Kerber et al.(2007)] {Kerber2007} Kerber, L.O., Santiago,
  B.X., \& Brocato, E. 2007, \aap, 462, 139
\bibitem[Kirby et al.(2016)] {Kirby2016} Kirby, E.N., Guhathakurta,
  P., Zhang, A.J., et al. 2016, \apj, 819, 135
\bibitem[Koch \& McWilliam(2008)] {KochMcWilliam2008} Koch, A. \&
  McWilliam, A. 2008, \aj, 135, 1551
\bibitem[Kraft \& Ivans(2003)] {KraftIvans2003} Kraft, R.P, \& Ivans,
  I.I. 2003, \pasp, 115, 143
\bibitem[Lambert \& Ries(1981)] {LambertRies1981} Lambert, D.L. \&
  Ries, L.M. 1981, \apj, 248, 228
\bibitem[Lapenna et al.(2012)] {Lapenna2012} Lapenna, E., Mucciarelli,
  A., Origlia, L., \& Ferraro, F.R. 2012, \apj, 761, 33
\bibitem[Lind et al.(2009)] {Lind2009} Lind, K., Asplund, M., \&
  Barklem, P.S. 2009, \aap, 503, 541
\bibitem[Lind et al.(2011)] {Lind2011} Lind, K., Asplund, M., Barklem,
P.S., \& Belyaev, A.K. 2011, \aap, 528, 103
\bibitem[Mackey \& Gilmore(2003)] {MackeyGilmore2003} Mackey, A.D. \&
  Gilmore, G.F. 2003, \mnras, 338, 85
\bibitem[Martell et al.(2016)] {Martell2016} Martell, S., Shetrone,
  M., Lucatello, S., et al. 2016, arXiv:1605.05792
\bibitem[Mashonkina et al.(2000)] {Mashonkina2000} Mashonkina, L.I.,
Shimanski\u{i}, V.V., \& Sakhibullin, N.A. 2000, \textit{Astronomy
Reports}, 44, 790
\bibitem[McCall(2004)] {McCall2004} McCall, M.L. 2004, \aj, 128, 2144
\bibitem[McWilliam et al.(1995)] {McWilliam1995} McWilliam, A.,
Preston, G.W., Sneden, C., \& Searle, L. 1995b, \aj, 109, 2757
\bibitem[McWilliam et al.(2013)] {McW2013} McWilliam, A., Wallerstein,
  G., \& Mottini, M. 2013, \apj, 778, 149
\bibitem[Meszaros et al.(2015)] {Meszaros2015} Meszaros, S., Martell,
  S.L., Shetrone, M., et al. 2015, \aj, 149, 153
\bibitem[Mucciarelli et al.(2008)] {Mucciarelli2008} Mucciarelli, A.,
Carretta, E., Origlia, L., \& Ferraro, F.R. 2008, \aj, 136, 375
\bibitem[Mucciarelli et al.(2009)] {Mucciarelli2009} Mucciarelli, A.,
  Origlia, L., Ferraro, F.R., \& Pancino, E. 2009, \apj, 695, 134
\bibitem[Mucciarelli et al.(2010)] {Mucciarelli2010} Mucciarelli, A.,
  Origlia, L., \& Ferraro, F.R. 2010, \apj, 717, 277
\bibitem[Mucciarelli et al.(2011)] {Mucciarelli2011} Mucciarelli, A.,
  Cristallo, S., Brocato, E., et al. 2011, \mnras, 413, 837
\bibitem[Mucciarelli et al.(2014a)] {Mucciarelli2014} Mucciarelli, A.,
  Dalessandro, E., Ferraro, F.R., Origlia, L., \& Lanzoni, B. 2014a,
  \apj, 793, 6
\bibitem[Mucciarelli et al.(2014b)] {Mucciarelli2014Li} Mucciarelli,
  A., Salaris, M., Bonifacio, P., Monaco, L., \& Villanova, S. 2014b,
  \mnras, 444, 1812
\bibitem[Niederhofer et al.(2016)] {Niederhofer2016} Niederhofer, F.,
  Bastian, N., Kozhurina-Platais, V., Hilker, M., de Mink, S.E.,
  Cabrera-Ziri, I., Li, C., \& Ercolano, B. 2016, \aap, 586, 148
\bibitem[Piatti et al.(2012)] {Piatti2012} Piatti, A.E., Geisler, D.,
  \& Mateluna, R. 2012, \aj, 144, 100
\bibitem[Pietrinferni et al.(2004)] {BaSTIref} Pietrinferni, A.,
Cassisi, S., Salaris, M. \& Castelli, F. 2004, \apj, 612, 168 
\bibitem[Pomp\'{e}ia et al.(2008)] {Pompeia2008} Pomp\'{e}ia, L.,
  Hill, V., Spite, M., et al. 2008, \aap, 480, 379
\bibitem[Pritzl et al.(2005)] {Pritzl2005} Pritzl, B.J., Venn, K.A.,
  \& Irwin,   M. 2005, \aj, 129, 2232
\bibitem[Qian \& Wasserburg(2008)] {QianWasserburg2008} Qian, Y.-Z. \&
  Wasserburg, G.J. 2008, \apj, 687, 272
\bibitem[Reddy et al.(2006)] {Reddy2006} Reddy, B.E., Lambert, D.L.,
\& Prieto, C.A. 2006, \mnras, 367, 1329
\bibitem[Sakari et al.(2011)] {Sakari2011} Sakari, C.M., Venn, K.A.,
  Irwin, M., Aoki, W., Arimoto, D., \& Dotter, A. 2011, \apj, 740, 106
\bibitem[Sakari et al.(2013)] {Sakari2013} Sakari, C.M., Shetrone, M.,
Venn, K., McWilliam, A., \& Dotter, A. 2013, \mnras, 434, 358
\bibitem[Sakari et al.(2014)] {Sakari2014} Sakari, C.M., Venn, K.,
Shetrone, M., Dotter, A., \& Mackey, D. 2014, \mnras, 443, 2285
\bibitem[Sakari et al.(2015)] {Sakari2015} Sakari, C.M., Venn, K.A.,
  Mackey, D., et al. 2015, \mnras, 448, 1314
\bibitem[Sakari et al.(2016)] {Sakari2016} Sakari, C.M., Shetrone,
  M.D., Schiavon, R.P., et al. 2016, \apj, {\it in press}
\bibitem[Schlafly \& Finkbeiner(2011)] {SchlaflyFinkbeiner2011}
  Schlafly, E.F. \& Finkbeiner, D.P. 2011, \apj, 737, 103
\bibitem[Shetrone et al.(2003)] {Shetrone2003} Shetrone, M., Venn, K.,
Tolstory, E., et al. 2003, \aj, 125, 684
\bibitem[Simmerer et al.(2004)] {Simmerer2004} Simmerer, J., Sneden,
  C., Cowan, J.J., Collier, J., Woolf, V.M., \& Lawler, J.E. 2004,
  \apj, 617, 1091
\bibitem[Skrutskie et al.(2006)] {2MASSref} Skrutskie, M.F., Cutri,
  R.M., Stiening, R., et al. 2006, \aj, 131, 1163
\bibitem[Smecker-Hane \& McWilliam(2002)] {SHMcW2002} Smecker-Hane,
  T.A. \& McWilliam, A. 2002, arXiv:0205411
\bibitem[Sneden(1973)]{Sneden} Sneden, C. 1973, \apj, 184, 839
\bibitem[Sneden et al.(1997)] {Sneden1997} Sneden, C., Kraft, R.P.,
Shetrone, M.D., Smith, G.H., Langer, G.E., \& Prosser, C.F. 1997, \aj,
114, 1964
\bibitem[Stetson \& Pancino(2008)] {DAOSPECref} Stetson, P.B. \&
Pancino, E. 2008, \pasp, 120, 1332
\bibitem[Tsujimoto \& Bekki(2012)] {TsujimotoBekki2012} Tsujimoto,
  T. \& Bekki, K. 2012, \apj, 751, 35
\bibitem[Tsujimoto \& Shigeyama(2014)] {TsujimotoShigeyama2014}
Tsujimoto, T. \& Shigeyama, T. 2014, \apjl, 795, L18
\bibitem[Van der Swaelmen et al.(2013)] {VanderSwaelmen2013} Van der
  Swaelmen, M., Hill, V., Primas, F., \& Cole, A.A. 2013, \aap, 560,
  A44
\bibitem[Venn et al.(2004)] {Venn2004} Venn, K.A., Irwin, M.,
Shetrone, M.D., et al. 2004, \aj, 128, 1177
\bibitem[Woosley \& Weaver(1995)] {WoosleyWeaver1995} Woosley, S.E. \&
  Weaver, T.A. 1995, \apjs, 101, 181
\bibitem[Yan et al.(2015)] {Yan2015} Yan, H.L., Shi, J.R., \& Zhao,
  G. 2015, \apj, 802, 36
\end{thebibliography}
\end{document}